\documentclass[%
 reprint,
 amsmath,amssymb,
 aps,
]{revtex4-2}

\usepackage{graphicx}
\usepackage{dcolumn}
\usepackage{bm}
\usepackage{xcolor}

\usepackage{textgreek} 
\usepackage{eucal}

\begin{document}

\preprint{APS/123-QED}

\title{To cross or not to cross: collective swimming of \emph{Escherichia coli} under two-dimensional confinement}

\author{Dipanjan Ghosh}
\thanks{ghosh135@umn.edu}
\affiliation{\mbox{Department of Chemical Engineering and Materials Science, University of Minnesota, Minneapolis, Minnesota 55455, USA}}

\author{Xiang Cheng}%
\thanks{xcheng@umn.edu}
\affiliation{\mbox{Department of Chemical Engineering and Materials Science, University of Minnesota, Minneapolis, Minnesota 55455, USA}}

\date{\today}

\begin{abstract}
Bacteria in bulk fluids swim collectively and display fascinating emergent dynamics. Although bacterial collective swimming in three-dimensional (3D) geometries has been well studied, its counterpart in confined two-dimensional (2D) geometries relevant to natural habitats of bacteria is still poorly understood. Here, through carefully designed experiments on \textit{Escherichia coli} in Hele-Shaw chambers, we show that a small change in the degree of confinement leads to a drastic change in bacterial collective swimming. While long-range nematic order emerges for bacteria that can cross during encounters, a slight decrease of the chamber height prevents the crossing, leading to the formation of bacterial clusters with short-range polar order. By tracking the swimming kinetics of individual bacteria, we reveal the microscopic origins of the two collective phases. Our study provides important insights into bacterial collective swimming under confinement and demonstrates a convenient way to control the emergent symmetry of collective phases.

\end{abstract}

\maketitle

\section{Introduction} 

Collective motion of bacteria epitomizes the emergent dynamics of active matter \cite{ramaswamy2010mechanics,marchetti2013hydrodynamics,schwarz2016review,beer2019review}, which leads to unusual transport properties of bacterial suspensions and confers upon bacteria evolutionary advantages crucial for their survival \cite{lopez2015superfluids,peng2016diffusion,guo2018shearbanding}. The natural habitats of bacteria often consist of confined spaces such as thin biofilms on solid substrates \cite{conrad2018confined}, pores of the soil \cite{ranjard2001quantitative,Bhattacharjee2019}, and the interstitial confines of tissues \cite{ki2008bacterial}. Consequently, understanding the collective dynamics of bacteria in confined systems is vital for deciphering various life-supporting activities of bacteria in their natural environments. However, although the collective dynamics of 3D bulk bacterial suspensions have been extensively studied in recent years \cite{Dombrowski2004collective,Sokolov2009collective,wensink2012meso,peng2021imaging,liu2020density}, our understanding of the dynamics of bacterial suspensions under geometric confinement is still primitive. 

Uncovering the dynamics of bacteria in confined geometries will provide fundamental insights into not only biological processes of practical importance but also the novel emergent collective behavior of active matter. While the long-range hydrodynamic interaction plays a leading role in inducing collective swimming in 3D bacterial suspensions \cite{peng2021imaging}, both the nature and strength of the interaction are strongly modified in confined systems. Particularly, for bacteria confined between two rigid walls, the far-field flow generated by a bacterium has the symmetry of a source dipole \cite{brotto2013hydrodynamics,tsang2014flagella}, qualitatively different from the force dipole flow in a 3D bulk fluid. More importantly, the short-range steric interaction that is inconsequential in 3D suspensions becomes essential in mediating the collective dynamics of bacteria in confined systems \cite{Aranson2007steric,baskaran2010steric}. These qualitative changes of inter-bacterial interactions result in novel collective phases of bacterial suspensions in confined systems \cite{nishiguchi2017long, swiecicki2013swimming,be2020phase}, which cannot persist in 3D bulk suspensions.

Inspired by different confined geometries in nature, several different types of confinement have been studied in experiments. Weak 3D confinement has been imposed by either narrow microfluidic channels or small droplets, where the chaotic turbulent-like flow induced by collective bacterial swimming in 3D is rectified into a persistent unidirectional flow \cite{wioland2013confinement,lushi2014confinement,wioland2016confinement,liu2019confinement}. Extensive studies have also been conducted for 2D confinement with a monolayer of bacteria swarming on agar substrates \cite{wu2015collective, kearns2010field,zhang2010collective,be2020phase}. Such a 2D geometry possesses a stress-free air-fluid interface, therefore relaxing the degree of confinement from the perspective of hydrodynamics. Recently, strong 2D confinement has also been implemented, where bacteria are confined in a Hele-Shaw cell with two rigid walls \cite{nishiguchi2017long,swiecicki2013swimming}. Using elongated swimming cells of \textit{Escherichia coli} (\textit{E. coli}) confined between two walls, Nishiguchi \textit{et al.} reported the emergence of collective bacterial swimming with long-range nematic order \cite{nishiguchi2017long}. In contrast, Swiecicki \textit{et al.} observed the formation of bacterial clusters with local polar order in a similar geometry \cite{swiecicki2013swimming}. Why do similar experiments yield collective phases with qualitatively different symmetries? How does confinement modify the inter-bacterial interactions and affect the collective swimming of bacterial suspensions? We aim to address these questions in our study.

Our study focuses on bacterial suspensions under strong 2D confinement in a Hele-Shaw cell. We find that the collective swimming of bacteria is sensitive to the degree of confinement. A small variation in the gap thickness between two rigid walls can trigger a drastic change of the emergent collective phase of bacteria and yield qualitatively different symmetries. The finding resolves the controversy surrounding the contradictory observations on bacterial dynamics under confinement from previous experiments. Our study further reveals that the emergence of the different collective phases is associated with the change of the microscopic inter-bacterial interaction. While bacteria that can cross over each other during close encounters form long-range nematic order, bacteria that are strictly constrained into a single layer under slightly stronger confinement assemble into transient clusters with local polar order. A subtle change in the microscopic inter-bacterial interaction has a profound effect on the emergent collective bacterial dynamics. Lastly, we show that the binary interaction between bacteria always favors nematic alignment, independent of the degree of confinement. Instead, the polar order of bacterial clusters arises from many-body steric interactions enabled by the non-crossing encounters between bacteria under strong confinement. These many-body interactions result in abnormally short swimming persistence and large velocity fluctuations of bacteria in the cluster phase. Taken together, our experiments on confined bacterial suspensions provide an excellent example illustrating the generic relation between the local particle interaction and the global symmetry of the emergent collective phases in active matter. Our study further demonstrates geometric confinement as an effective tool to control the collective dynamics of bacterial suspensions, paving a way to engineer the swimming behaviors of bacteria for practical applications.

\section{Experiment}

In our experiments, we use genetically modified light-powered {\it E. coli} (see Appendix \ref{app:methods}: Methods), whose average swimming velocity $V$ can be controlled between 4 and 15 {\textmu}m/s by varying the intensity of incident light \cite{peng2021imaging,liu2020density}. In addition to bacterial swimming velocity, we also vary 2D bacterial number density $n$ between $1.3 \times 10^6$ up to $2.6 \times 10^7$ mm$^{-2}$. Above $2.6 \times 10^7$ mm$^{-2}$, bacteria are immotile in our confined cell, possibly due to the intertwining of flagellar bundles at high densities. The area fraction of the suspension is given by $\phi = nA$, where $A = 2.3$ {\textmu}m$^2$ is the cross-section area of bacteria in the 2D plane. We confine a suspension of {\it E. coli} of controlled volume in a Hele-Shaw cell made of a glass slide and a coverslip (see Appendix \ref{app:methods}: Methods). The lateral dimension of the cell is fixed at 18 mm by 18 mm, whereas the gap thickness of the cell is controlled by the volume of the suspension. We test two different suspension volumes, i.e., 0.7 {\textmu}L and 0.9 {\textmu}L, in our experiments. As the suspension is completely confined underneath the coverslip by capillary forces, the gap thickness is fixed at $h \approx 2.2$ {\textmu}m for the small-volume suspension and $h \approx 2.8$ {\textmu}m for the large-volume suspension. The cell is finally sealed on all sides by a UV-curable adhesive, which eliminates the influence of ambient airflow on the bacterial motion. Swimming bacteria in the cell are then imaged using an inverted microscope at a frame rate of 30 fps with a field of view of 232 {\textmu}m by 208 {\textmu}m. By post-processing the resulting images (see Appendix \ref{app:methods}: Methods), we identify both the position $\mathbf{r}$ and orientation $\theta$ of the bacterial body along the direction of swimming, and the instantaneous velocity $\mathbf{v}$ of individual bacteria.

\begin{figure*}

\centering
\includegraphics[width=\linewidth]{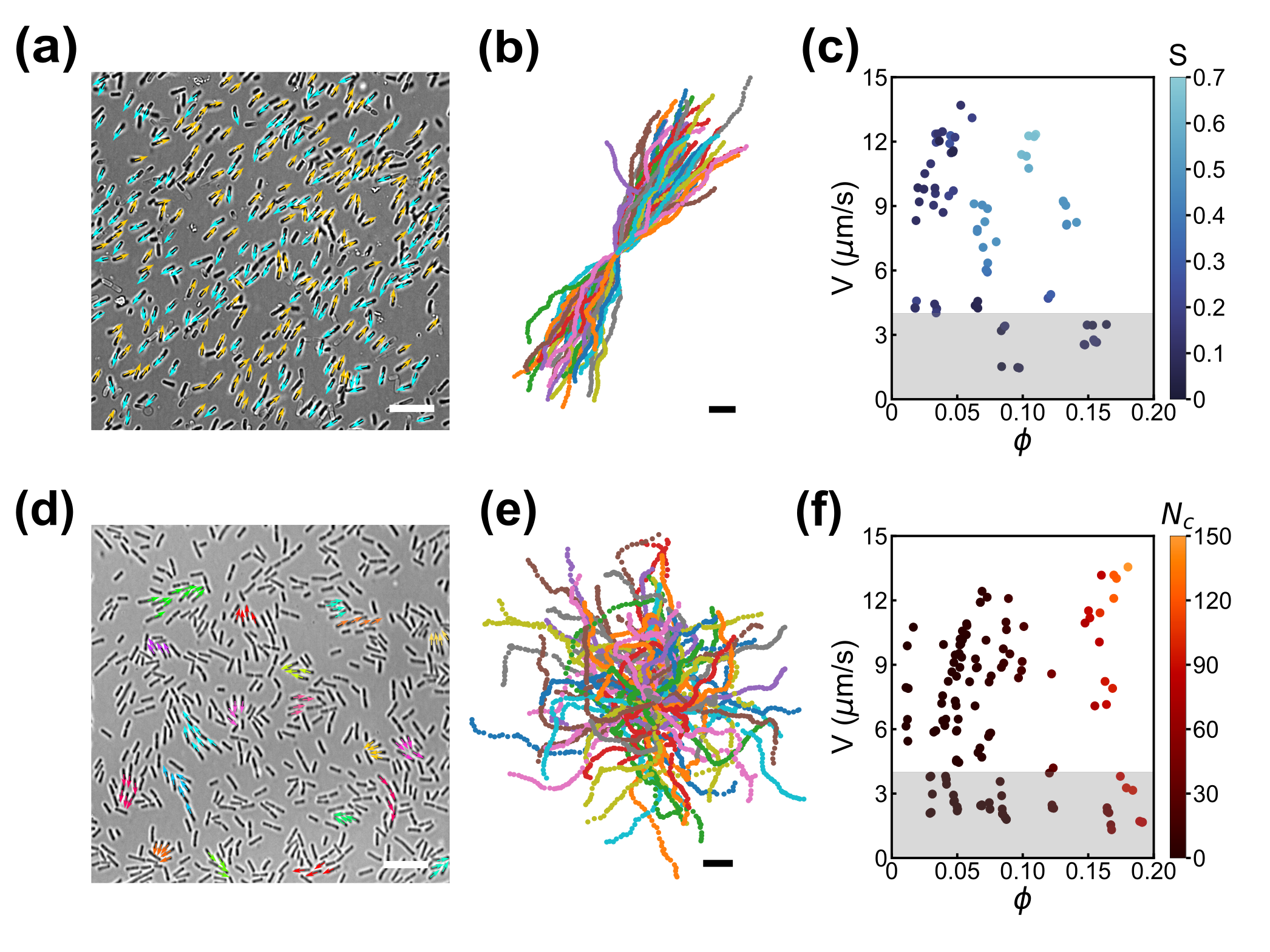}
\caption{Collective swimming of \textit{E. coli} in the quasi-2D (top row) and 2D (bottom row) geometries. (a) Microscopy image of a bacterial suspension in the quasi-2D geometry exhibiting long-range nematic order. The gold and cyan arrows represent the two directions along which they are predominantly oriented. (b) Representative trajectories of bacteria in the quasi-2D geometry, with the starting point of each trajectory translated to coincide to a single point at the center. The preferred orientation of the trajectories demonstrates the existence of the nematic order. (c) Phase diagram showing the dependence of the nematic order parameter $S$ on the bacterial swimming velocity $V$ and area fraction $\phi$. (d) Microscopy image of a bacterial suspension in the 2D geometry exhibiting bacterial clusters with short-range polar order. Arrows of different colors are used to mark different clusters. (e) Representative trajectories of bacteria in the 2D geometry, with the starting point of each trajectory translated to coincide to a single point at the center. (f) Phase diagram showing the dependence of the number of clusters $N_c$ in the field of view on $V$ and $\phi$. The gray regions in (c) and (f) correspond to slow and randomly moving bacteria with velocities below 4 {\textmu}m/s, observed at low light intensities.  Scale bars are 10 {\textmu}m.}

\label{fig:fig1}
\end{figure*}

\section{Results}

\subsection{Collective swimming in the 2D and quasi-2D geometries}

With the small change of the gap thickness, we observe two qualitatively different collective phases of bacterial suspensions, which are characterized by different orientational orders of bacteria. When the gap thickness is large, a case we shall refer to as the quasi-2D geometry below, bacteria show random swimming at low densities $\phi$ and small bacterial swimming velocities $V$. With increasing $\phi$ and $V$, bacteria tend to align nematically over a long range. At high $\phi$ and $V$, the collective motion of bacteria shows a clear long-range nematic order, where bacterial bodies align along a preferred direction with bacteria themselves swimming either parallel or antiparallel along the direction (Fig.~\ref{fig:fig1}(a),  Fig.~\ref{fig:fig1}(b), and Supplementary Movie 1 \cite{ghosh2021supplementary}). We quantify the strength of the nematic alignment using the order parameter $S = \sqrt{(\overline{\cos2\theta})^2 + (\overline{\sin2\theta})^2}$ (Fig.~\ref{fig:fig1}(c)), where the overbars indicate averages taken over all the bacteria in the field of view. $S = 1$ indicates a perfect alignment, whereas $S = 0$ for random orientations. Consistent with our direct observation, $S$ increases with $\phi$ and $V$ and reaches $S = 0.7$ at high $\phi$ and $V$. The quasi-long-range nematic phase has been reported in experiments with filamentous cells of \textit{E. coli} with a large aspect ratio of the bacterial body at 25 \cite{nishiguchi2017long}. Here, we demonstrate that the long-range nematic order can also arise in bacteria with the aspect ratio of wild-type \textit{E. coli} at 3.5 under strong confinement.

For the small gap thickness, a case we shall refer to as the 2D geometry below, bacteria also show random swimming at low $\phi$ and small $V$, similar to those in the quasi-2D geometry. Nevertheless, with increasing $\phi$ and $V$, instead of the long-range nematic order, bacteria form transient clusters with short-range polar order (Fig.~\ref{fig:fig1}(d) and Supplementary Movie 2 \cite{ghosh2021supplementary}). Such structures have been termed as ``bacterial rafts'' by Swiecicki \emph{et al} \cite{swiecicki2013swimming}. The trajectories of individual bacteria in 2D are much less persistent than those in quasi-2D, without a clear sign of the nematic order (Fig.~\ref{fig:fig1}(e)). To characterize the collective behavior in the 2D geometry, we assign bacteria into clusters based on the distance between them and the difference between their orientations. A pair of bacteria are adjacent neighbors when the distance between the centroids of their bodies $\Delta r < 3$ {\textmu}m and the difference between their body orientation $\Delta \theta < 30 ^{\circ}$. A cluster is then defined as a group of bacteria where each bacterium belonging to the group is an adjacent neighbor with at least one other bacterium from the same group. Furthermore, a cluster must consist of at least 4 bacteria. We quantify the extent of cluster formation by counting the number of bacterial clusters $N_c$ in our field of view. $N_c$ increases with both $\phi$ and $V$ and reaches $N_c = 150$ at high $\phi$ and $V$ (Fig.~\ref{fig:fig1}(f)).

\subsection{Binary collisions favor weak nematic alignment in both geometries}

\begin{figure*}
\centering
\includegraphics[width=\linewidth]{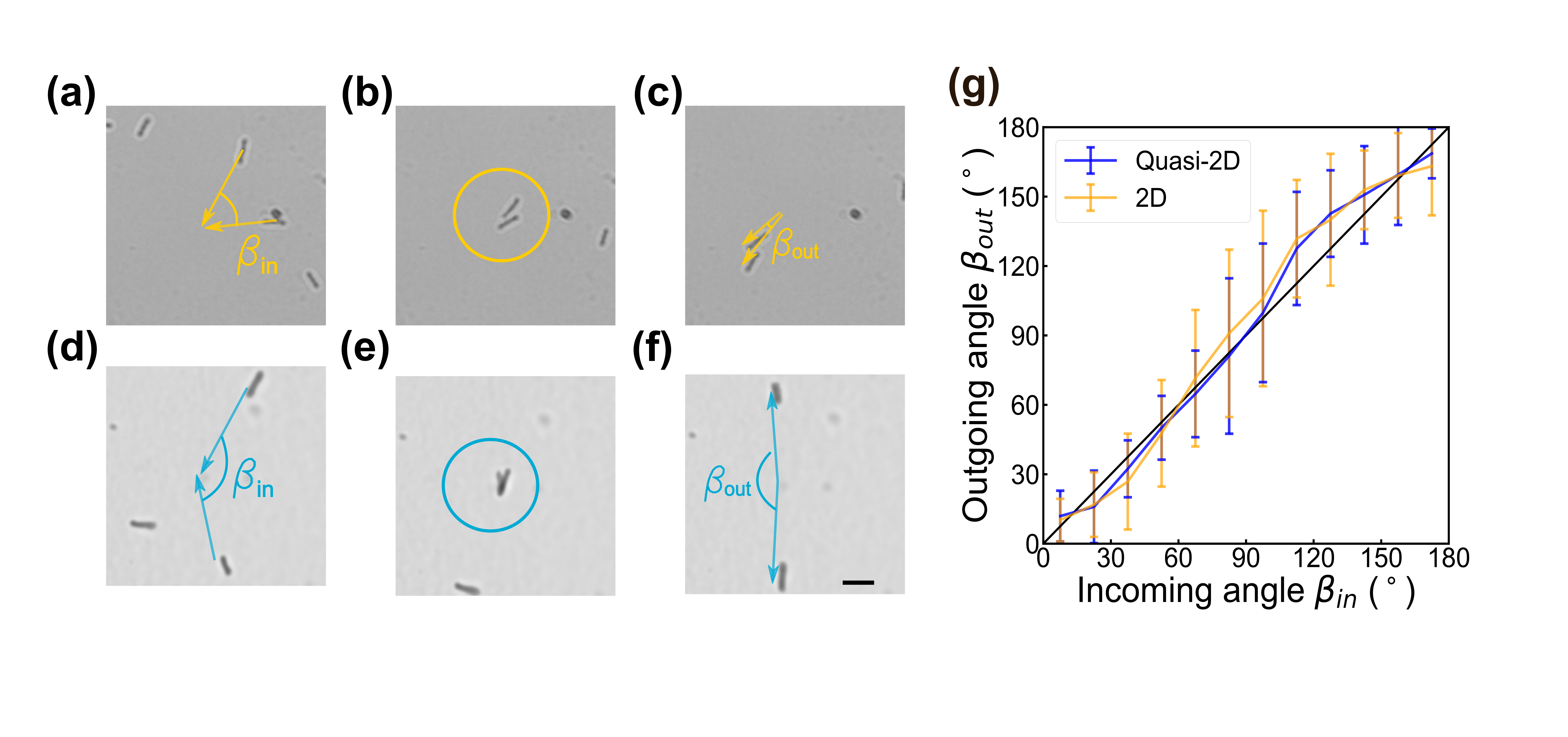}
\caption{Nematic alignment induced by binary collisions. (a)-(c) show the time-lapse frames of a collision of two bacteria in the 2D geometry with an acute incoming angle at $t = -1$, 0 and 1 s. (d)-(f) show the time-lapse frames of a collision of two bacteria in the quasi-2D geometry with an obtuse incoming angle at $t = -1$, 0 and 1 s. Arrows indicate the direction of bacterial swimming. Scale bar is 5 {\textmu}m. (g) The outgoing angle $\beta_{out}$ versus the incoming angle $\beta_{in}$ for the quasi-2D (blue) and 2D (gold) geometries. A total of 1,119 collisions for quasi-2D geometry and a total of 1,287 collisions for the 2D geometry have been considered. The data is binned into $15^{\circ}$ intervals of $\beta_{in}$. The blue and gold lines connect the means of $\beta_{out}$ for these intervals, and the errorbars indicate standard deviations. The black straight line represents the condition $\beta_{in} = \beta_{out}$, corresponding to a collision without any alignment.}
\label{fig:fig2}
\end{figure*}

Why does a small change in the gap thickness lead to such a drastic change in the collective dynamics of bacteria? To answer this question, we first examine the pairwise interaction between bacteria at low $\phi$ and large $V$ in both the quasi-2D and 2D geometries. Specifically, we analyze experiments in the dilute limit with $\phi = 0.05$ and high bacterial activity $V = 12$ {\textmu}m/s. A collision event between a pair of bacteria is defined when the distance between the centroids of their body $\Delta{r} = |\mathbf{r_1}-\mathbf{r_2}| < 3$ {\textmu}m, where $\mathbf{r_1}$ and $\mathbf{r_2}$ are the centroids of the two bacteria, respectively. The positions of the bacterial bodies are then tracked starting from 1 second before the collision event to 1 second after the collision event. Bacteria typically swim a distance of at least one cell-body length from their point of collision in 1 second, which is sufficient to determine their new direction of motion after a collision. The angle subtended by the positions of the two bacteria before the collision is termed the incoming angle $\beta_{in}$, whereas the angle subtended by their positions after the collision is called the outgoing angle $\beta_{out}$ (Fig.~\ref{fig:fig2}[(a)-(f)]).

Figure~\ref{fig:fig2}(g) shows $\beta_{out}$ as a function of $\beta_{in}$ for the quasi-2D and 2D geometries. Surprisingly, even though the emergent collective phases are qualitatively different, the binary interactions between bacteria are quantitatively similar in the two geometries. For acute incoming angles, the outgoing angle is slightly decreased, indicating a tendency for weak polar alignment, whereas, for obtuse incoming angles, a slight increase in the outgoing angle captures weak anti-polar alignment. Thus, the binary interactions in both the quasi-2D and 2D geometries indicate a weak nematic symmetry of pairwise interactions \cite{nishiguchi2017long,tanida2020gliding,sciortino2021pattern,grossmann2020particle}, without any discernible bias towards polar alignment \cite{denk2020pattern}. In the quasi-2D geometry, at high densities at which the nematic order is observed, a bacterium undergoes multiple successive binary collisions with its neighbors. Even though a single collision imparts only weak nematic alignment, the collective alignment resulting from multiple collisions is sufficient to induce long-range nematic order in the quasi-2D geometry. However, the difference in the collective behaviors in the two geometries, particularly, the rise of bacterial clusters with local polar order in 2D, cannot be explained by binary collisions.

\subsection{To cross or not to cross}

A detailed examination of bacterial dynamics at both low and high $\phi$ reveals a key difference in bacterial interactions in the quasi-2D and 2D geometries. While bacteria can cross over each other during a collision in the quasi-2D geometry (Fig.~\ref{fig:fig2}(e)), we do not observe any bacterial crossing in the 2D geometry (Fig.~\ref{fig:fig2}(b)). The tighter confinement of the 2D geometry strictly constrains bacteria in a single layer. Our experiments thus suggest that decreasing the thickness of the Hele-Shaw cell, thereby switching off the ability of bacteria to cross over, drastically alters their emergent collective swimming behaviors. 

\begin{figure*}
\centering
\includegraphics[width=\linewidth]{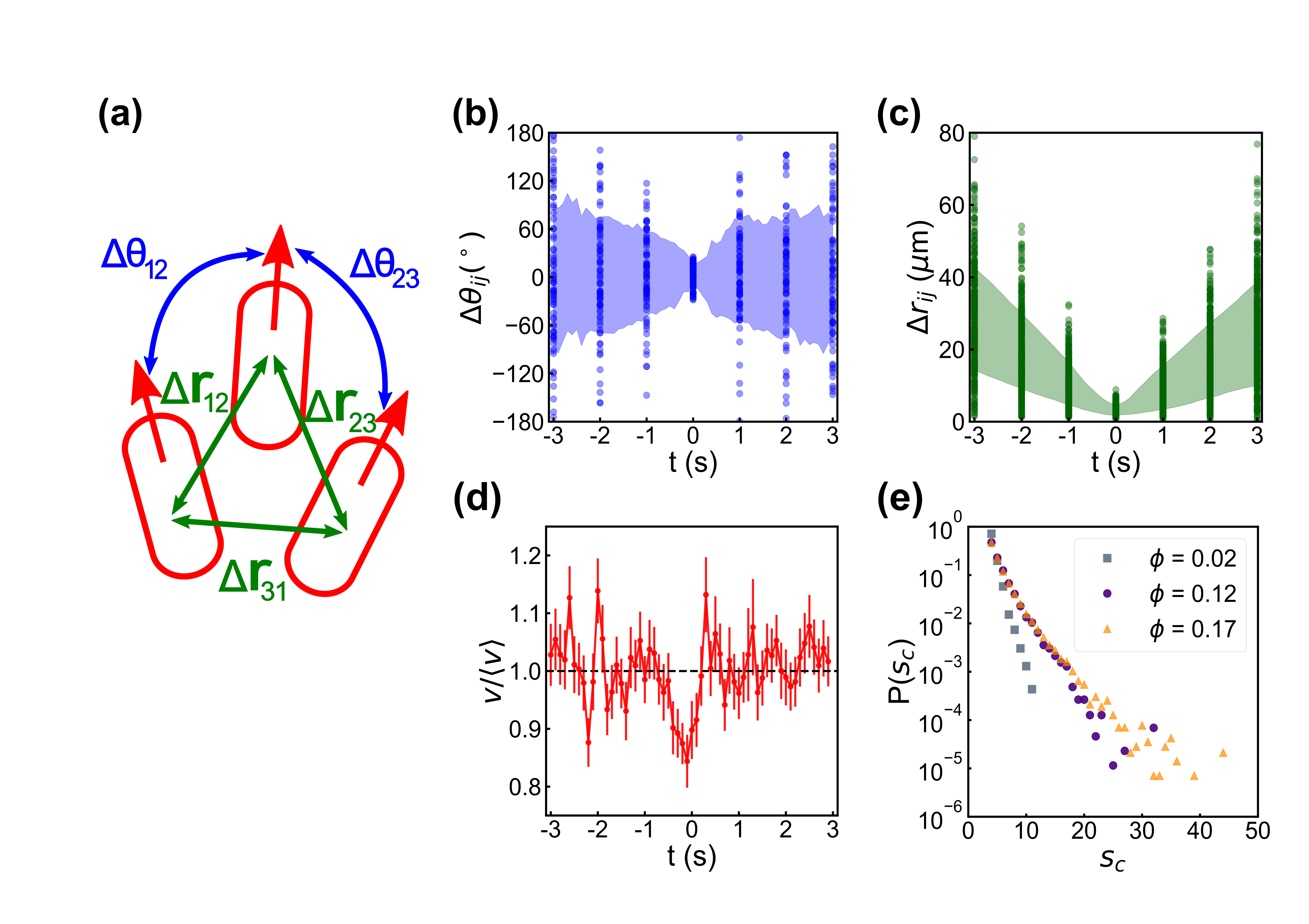}
\caption{Transient dynamics of bacterial clusters in the 2D geometry. (a) A schematic showing the adjacent angles $\Delta \theta_{ij}$ and the pairwise distance $\Delta r_{ij}$ of three bacteria. (b) and (c) The time evolution of $\Delta \theta_{ij}$ and $\Delta r_{ij}$. Disks show the data, whereas the shaded region indicates one standard deviation around the mean of the data. To avoid crowding, we only show the data at the integer time of $t=-3$, $-2$, $-1$, 0, 1, 2 and 3 s. The measurements are conducted with the time resolution of $\Delta t = 0.1$ s. (d) The time evolution of the normalized velocity of bacteria $v / \langle v \rangle$. The error bars represents the standard errors of the normalized velocity of all bacteria in clusters. The black dashed line corresponds to the mean $v/\langle v \rangle = 1$. (e) The probability distribution $P(s_c)$ of the size of clusters $s_c$ in the 2D geometry at three different area fractions $\phi=0.02$ (gray squares), $\phi = 0.12$ (purple circles) and $\phi=0.17$ (gold triangles).} 

\label{fig:fig3}
\end{figure*}

The relation between the emergent order of collective phases and the ability of individual active particles to cross during collisions has also been observed in other 2D active matter systems. In numerical simulations of active rods, the crossover between two particles can be controlled by the strength of the repulsive interparticle potential and the self-propulsion speed of active particles \cite{bar2020self}. For soft potentials at high speeds, a condition where particles can overlap and cross over each other, the symmetry of the emergent collective phase is nematic \cite{ginelli2010large, grossmann2020particle, shi2018self}. In contrast, slower speeds and stiffer repulsive potentials lead to non-crossing inter-bacterial interaction and give rise to polar clusters \cite{peruani2006nonequilibrium,grossmann2020particle,shi2018self} and bands \cite{abkenar2013collective,weitz2015self}. A recent experiment with a 2D motility assay of microtubules has also shown that a nematic order emerges when the microtubules are able to cross over each other, whereas polar clusters form when they are unable to cross \cite{tanida2020gliding}. There, the ability of microtubules to cross was controlled by the density of motor proteins fixed on substrates. A low motor-protein density gives more flexibility to the tip of a microtubule and allows it to climb over other microtubules during a collision. Thus, in combination with these previous numerical and experimental findings, our experiments with swimming bacteria---a premier example of active matter---provide strong evidence illustrating a universal feature of 2D active matter: the ability of active particles to cross dictates the symmetry of emergent collective phases. Rather than modifying the properties of individual active particles, our study demonstrates that geometric confinement can be used as a simple and convenient tool to control the crossing ability of active particles and manipulate the collective dynamics of 2D active matter.

\subsection{The rise and fall of bacterial clusters}

While the long-range nematic order of bacteria in the quasi-2D geometry has already been shown to originate from the binary collision of bacteria in the dilute limit (Fig.~\ref{fig:fig2}(g)), bacterial clusters with the local polar order in the 2D geometry must arise from the many-body interactions enabled by the non-crossing collision at high densities \cite{suzuki2015polar,sciortino2021pattern}.

To understand the origin of bacterial clusters, we image the dynamic process of cluster formation in the 2D geometry. Specifically, we analyze 17 representative bacterial clusters at $\phi = 0.15$. Each cluster contains 4-9 bacteria, giving a total of 117 bacteria across all of the clusters. The time at which they are identified is assigned as the reference time $t = 0$. The positions and body orientations of bacteria in these clusters are then tracked from $t = -3$ to 3 s at 0.1 s intervals. At each time step, we calculate the difference between the angles of adjacent neighbors, $\Delta \theta_{ij}$, as well as the pairwise distances between all the members of a bacterial cluster, $\Delta r_{ij}$ (see the schematic in Fig.~\ref{fig:fig3}(a) for the definition). Figures~\ref{fig:fig3}(b) and ~\ref{fig:fig3}(c) show $\Delta \theta_{ij}$ of all adjacent pairs of bacteria and $\Delta r_{ij}$ of all bacterial pairs in the 17 bacterial clusters as a function of time. From $t = -3$ to 0 s, both the extents of $\Delta \theta_{ij}$ and $\Delta r_{ij}$ decrease, indicating bacteria coming together and aligning to form a cluster. Subsequently, $\Delta \theta_{ij}$ and $\Delta r_{ij}$ increase with time from $t = 0$ to 3 s, showing the gradual dissolution of the clusters over time and revealing the transient nature of bacterial clusters in the 2D geometry.

Along with $\Delta \theta_{ij}$ and $\Delta r_{ij}$, we also measure the velocity of bacteria in the process of cluster formation and dissolution. 
We define $v/\langle v \rangle$ as the normalized average velocity of an individual bacterium, where $v$ is the instantaneous velocity of the bacterium and $\langle v \rangle$ is the time-averaged velocity of that particular bacterium. Figure~\ref{fig:fig3}(d) shows the time evolution of $v/\langle v \rangle$ of bacteria in clusters. Around $t = 0$, $v/\langle v \rangle$ decreases substantially about $15\%$ below its temporal average, suggesting an instantaneous slowing down at the instant of cluster formation. Here, the forward motion of a bacterium can be partially blocked by neighboring bacteria acting as obstacles \cite{yang2010swarm,sciortino2021pattern}. If multiple bacteria encounter the same obstacle, their velocities slow down instantaneously, which leads to the formation of a bacterial cluster. Each bacterium in the cluster further aligns with its neighbors, giving rise to local polar order \cite{grossmann2020particle,sciortino2021pattern}. If the bacterium aligned anti-parallel with its neighbors, it would simply slide away without joining the cluster. Since a bacterium can cross past its neighbors in the third dimension, the collision-induced slowdown---the key feature underlying the cluster formation---does not occur in the quasi-2D geometry. As the mechanism of collision-induced slowdown requires the presence of multiple neighbors, the clustering does not occur at low $\phi$ either. Even though the two bacteria undergoing a collision slow down temporarily, there is an insufficient number of neighboring bacteria at low $\phi$ that can join the pair before they separate.

The increase in adjacent angle differences (Fig.~\ref{fig:fig3}(b)) and in pairwise distances (Fig.~\ref{fig:fig3}(c)) after $t = 0$ suggest that the bacterial clusters in the 2D geometry are  transient. Due to their short lifetimes, clusters are unable to grow and remain small in size. We verify this by measuring the sizes $s_c$ of the observed clusters at different densities for a high bacterial velocity of $V = 12$ {\textmu}m/s. At large $\phi$, the probability distribution of cluster sizes $P(s_c)$ is independent of $\phi$. The maximum cluster size is  $\sim \CMcal{O}(10)$. In comparison, the number of bacteria in the field of view is on the order of $10^3$. Thus, unlike the long-range nematic order that is formed by all bacteria in the field of view, the polar clusters are short-range and spatially localized.

Bacteria in the 2D geometry form clusters as their forward motion is being partially blocked by a common obstacle. When this obstacle moves away, the velocities of the bacteria in the cluster increase, as indicated by Fig.~\ref{fig:fig3}(d) after $t = 0$. Figure~\ref{fig:fig3}(c) further shows that the pairwise distances between the bacteria in a cluster increase after $t = 0$, implying that the members of the cluster are moving apart. Such a mechanism is possible if the swimming speed of a bacterium in a cluster depends on its relative position in the cluster. Specifically, a bacterium that does not have any neighbors ahead of it swims faster and moves away from the cluster. This difference in relative velocities between the bacteria in a cluster is ultimately responsible for the transient nature and the small size of bacterial clusters.

\begin{figure*}
\centering
\includegraphics[width=\linewidth]{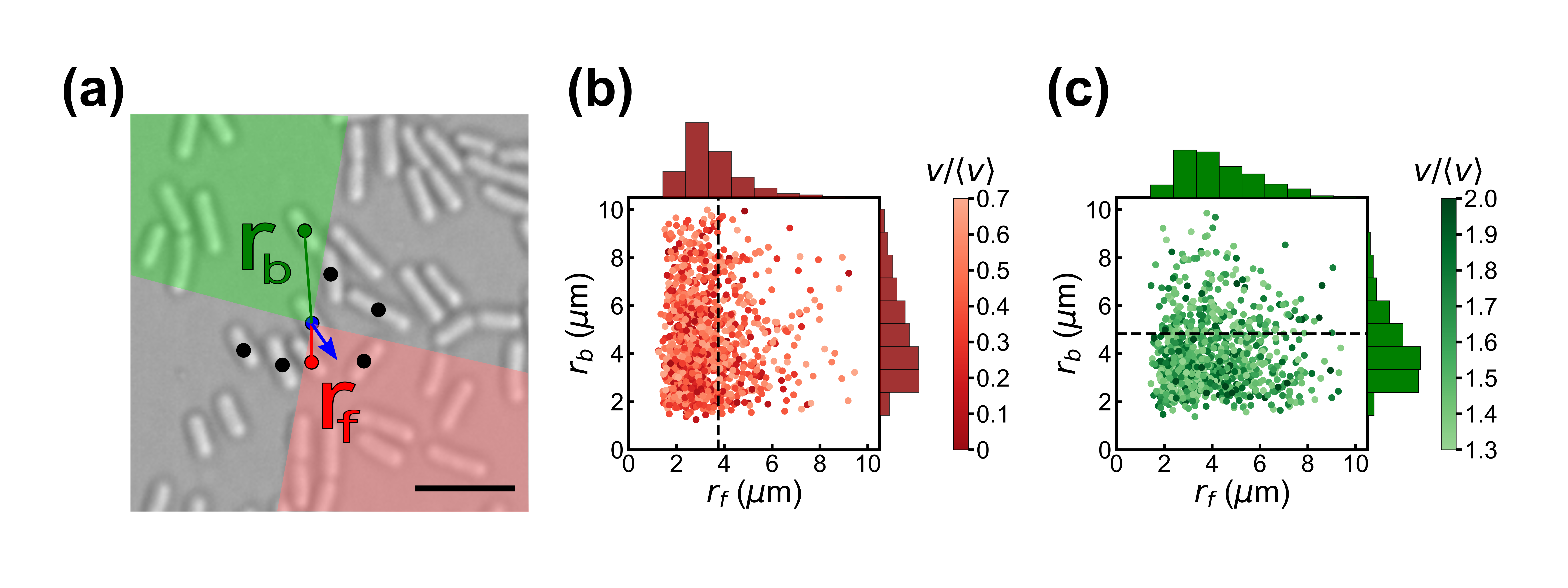}
\caption{Dependence of bacterial velocity on the relative positions of their neighbors in the 2D geometry. (a) The positions of the nearest neighbors of the reference bacterium marked by the blue arrow. The black, green and red dots denote the Voronoi nearest neighbors of the reference bacterium. Nearest neighbors are searched in the forward (pink shaded) region and the backward (green shaded) region, where each region lies within $\pm 45 ^{\circ}$ of the direction parallel and anti-parallel to the direction of propulsion. The distance to the closest nearest neighbor in the forward region (a red dot) is $r_f$ and the distance to the closest nearest neighbor in the backward region (a green dot) is $r_b$. (b) The distribution of $r_f$ and $r_b$ when bacterial velocity $v/ \langle v \rangle < 0.7$. (c) The distribution of $r_f$ and $r_b$ when bacterial velocity $v/ \langle v \rangle > 1.3$. The dashed black lines denote the 75th percentile of $r_f$ in (b) and the 75th percentile of $r_b$ in (c).} 

\label{fig:fig4}
\end{figure*}

Thus, the transient nature of bacterial clusters suggests a strong dependence of the swimming velocity of bacteria on the local bacterial density as well as the relative positions of their neighbors. To probe this dependence, we consider the velocity of bacteria with respect to the distances to their closest neighbors. Specifically, at each time step, we find the positions of the nearest Voronoi neighbors of the bacterium under consideration and calculate the distances between the bacterium and its nearest neighbors. We distinguish two types of neighbors. Neighbors lying within $45^{\circ}$ of the swimming direction of the bacterium are identified as forward neighbors, whereas neighbors lying within $45^{\circ}$ of the opposite direction of the swimming are identified as backward neighbors (Fig.~\ref{fig:fig4}(a)). The distance between the tracked bacterium and its forward nearest neighbor is termed $r_f$ and the distance to its backward nearest neighbor is termed $r_b$. We examine the dependence of the normalized velocity of the bacterium $v/\langle v \rangle$ on $r_f$ and $r_b$ for the cases leading to large velocity fluctuations, where $v/\langle v \rangle$ is at least $30\%$ above or below one. Figures~\ref{fig:fig4}(b) and ~\ref{fig:fig4}(c) show the joint plot of $v/\langle v \rangle < 0.7$ and $v/\langle v \rangle > 1.3$ as a function of $r_f$ and $r_b$, respectively. Most data with $v/\langle v \rangle < 0.7$ cluster around lower values of $r_f$, where there is a forward neighbor at a short distance (Fig.~\ref{fig:fig4}(b)). Quantitatively, $75\%$ of all the data points with $v/\langle v \rangle < 0.7$ have $r_f < 3.7$ {\textmu}m. Thus, the presence of a forward neighbor at a short distance in the path of the bacterium reduces its swimming velocity. This observation again confirms the collision-induced slowdown essential for the formation of bacterial clusters. More interestingly, most data with $v/\langle v \rangle > 1.3$ cluster around lower values of $r_b$ with $75\%$ of all data points having $r_b < 4.8$ {\textmu}m (Fig.~\ref{fig:fig4}(c)). This result suggests that the presence of a backward neighbor at a short distance behind a bacterium enhances its swimming speed. Such an enhancement promotes the quick dissolution of bacterial clusters after the removal of blockage. Bacteria in the front of a cluster accelerate to leave the cluster, giving rise to small transient bacterial clusters. The velocity enhancement due to backward neighbors is unique to swimming bacteria in confinement and has not been observed in experiments with swarming bacteria without confinement \cite{zhang2010collective} or clustering microtubules \cite{tanida2020gliding,sciortino2021pattern}.

Why does a backward neighbor enhance the swimming speed of bacteria? \textit{E. coli} are flagellated bacteria that swim due to the thrust force generated by the rotation of bacterial flagella \cite{vizsnyiczai2020transition}. We hypothesize that the presence of a neighboring cell body in a tightly confined geometry close to the flagella increases the thrust force, which causes an increase in the swimming speed.  Thus, the interactions with its forward and backward neighbors strongly affect the swimming velocity of the bacterium in the 2D geometry, which in turn affects the structure of bacterial clusters. Each bacterium in a cluster has a different distribution of forward and backward nearest neighbors, resulting in a large difference in the swimming velocities of bacteria in the cluster. Thus, bacteria from the same cluster swim at different speeds, leading to the quick dissolution of the cluster.

\subsection{Distinct single bacterial dynamics in the 2D and quasi-2D geometries}

The qualitative difference in the emergent collective phases in the 2D and quasi-2D geometries also implies a drastic difference in the swimming behavior of individual bacteria in these two geometries even at the same concentration. To highlight these differences, we compare the swimming trajectories in the 2D and quasi-2D geometries at a high density of $\phi = 0.15$ and a high swimming velocity of $V = 12$ {\textmu}m/s.

\begin{figure}[ht]
\centering
\includegraphics[width=\linewidth]{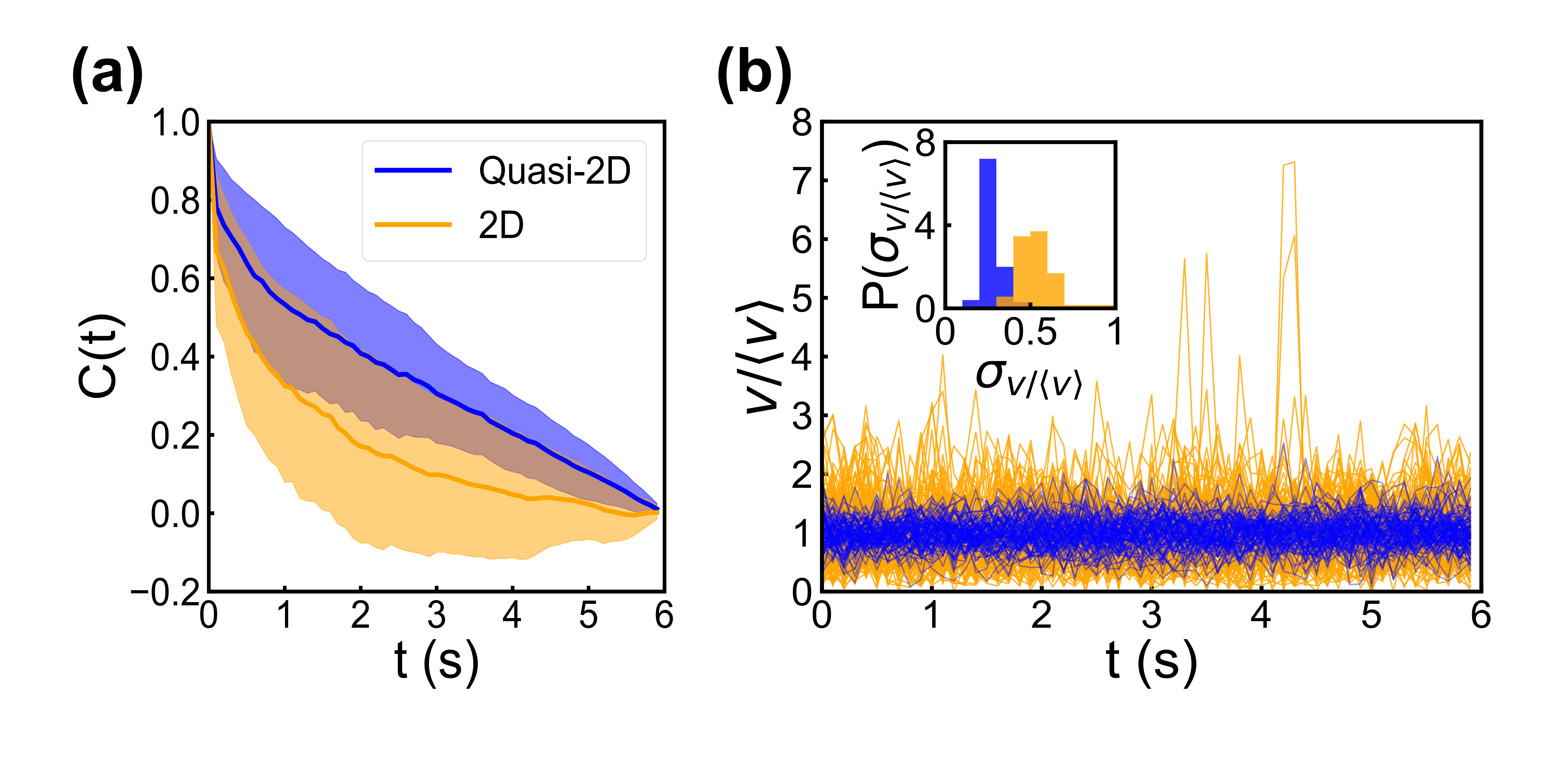}
\caption{Comparison of the swimming behaviors of bacteria in the 2D and quasi-2D geometries at the same bacterial area fraction of $\phi$ = 0.15. (a) The autocorrelation of the swimming direction of bacteria, $C(t)$, for the 2D (gold) and quasi-2D (blue) geometries. The thick lines are the means calculated over 100 bacteria in the quasi-2D geometry and over 117 bacteria in the 2D geometry, whereas the shaded region represents one standard deviation around the means. (b) The temporal evolution of the normalized velocity of individual bacteria, $v/ \langle v \rangle$, for the 2D (gold) and quasi-2D (blue) geometries. The inset shows the probability distribution of the standard deviation of $v/ \langle v \rangle$ of each bacterial trajectory, $P(\sigma_{v/\langle v \rangle})$, which quantify the magnitude of the fluctuations in the two geometries.}
\label{fig:fig5}
\end{figure}

First, we compute the persistence of the swimming direction of individual bacteria in the two geometries. The persistence of the swimming direction of bacteria can be quantified by the autocorrelation $C(t) = \langle \cos(\alpha(t_0)) \cos(\alpha(t_0+t)) \rangle_{t_0}$, where $\alpha(t)$ is the angle of the direction of the bacterial swimming velocity at time $t$ with respect to the $x$ axis in the lab frame. $C(t)$ decays faster in the 2D geometry compared to the quasi-2D geometry, implying a shorter persistence of the swimming bacteria in 2D than in quasi-2D (Fig.~\ref{fig:fig5}(a)). This short persistence in 2D is a direct consequence of the non-crossing collisions and the many-body steric interactions. As bacteria are unable to cross over during collision, they must change their swimming directions frequently at high $\phi$. In contrast, a bacterium in quasi-2D is able to cross over during collisions and maintain its swimming direction.

Second, we examine the temporal fluctuations of the velocity of individual bacteria in both geometries. Figure~\ref{fig:fig5}(b) shows the temporal variation of the normalized bacterial velocity $v(t)/\langle v \rangle$ in the 2D and quasi-2D geometries. The velocity fluctuations in the 2D geometry are significantly stronger than those in the quasi-2D geometry. The velocity fluctuation of each bacteria can be quantified by the standard deviation of its normalized velocity, $\sigma_{v/ \langle v \rangle}$ around the mean. Figure.~\ref{fig:fig5}(b) inset shows the probability distribution of $\sigma_{v/ \langle v \rangle}$ of all bacteria in the two different geometries. The average standard deviation of velocities is about 50\% of the mean velocity in the 2D geometry, whereas it is only 25\% of the mean in the quasi-2D geometry. This difference in velocity fluctuations can also be inferred from the strong many-body steric and near-field hydrodynamic interactions between the bacteria in the 2D geometry. A bacterium in 2D slows down instantaneously upon encountering a forward neighbor in its path and speeds up due to the presence of a backward neighbor close to it. At high densities, collisions between bacteria result in frequent speedups and slowdowns, contributing to the enhanced velocity fluctuations in the 2D geometry. In contrast, a bacterium in the quasi-2D geometry is able to slide past a neighbor in its path, without significantly slowing down or speeding up, resulting in weak velocity fluctuations.

\section{Discussion}

Our observations with confined swimming bacteria such as the existence of the nematic and cluster phase and the correlation of these phases with the crossing/non-crossing ability of individual units show striking resemblances to the collective dynamics of microtubules driven by motor proteins \cite{tanida2020gliding}. This finding suggests a universality of the collective behavior in the two arguably most important examples of biological active matter \cite{marchetti2013hydrodynamics}. Nevertheless, qualitative differences still exist due to the different natures of the two systems. Firstly, the propulsion mechanism of \textit{E. coli} via rotating flagella gives rise to non-trivial near-field hydrodynamic effects, which, when modified by geometric confinement, results in the enhanced velocity fluctuations and transient nature of bacterial clusters. In comparison, microtubule clusters at high densities are stable and long-lasting \cite{suzuki2017emergence}, because of the absence of destabilizing hydrodynamic interactions between microtubules. Stable microtubule clusters dynamically merge together into larger clusters or split into smaller clusters \cite{tanida2020gliding,sciortino2021pattern,suzuki2017emergence}. Such dynamic processes are not observed in bacterial clusters in our experiments due to the short lifetimes of the clusters.

Secondly, the collective phases of confined bacterial suspensions are uniform and stable throughout our experiments. We do not observe the coexistence or the dynamic switching between the nematic and polar phases, which have been reported in experiments of microtubules \cite{huber2018emergence}. In the microtubule system, the addition of a depleting agent increases the bonding between microtubule filaments and motor proteins attached to a substrate. The enhanced bonding to the substrate reduces the ability of the filaments to cross during a collision and therefore promotes the nematic alignment of microtubules. Consequently, nematic and polar collective phases are observed for high and low concentrations of the depleting agent, respectively. At intermediate concentrations of the depleting agent, the crossing and non-crossing of filaments are not well controlled, which leads to the coexistence of both nematic and polar phases in the microtubule system \cite{huber2018emergence}. In comparison, the ability of bacteria to cross is fixed globally by the gap thickness of confined geometries in our experiments. At a given gap thickness, the collective swimming behavior of bacteria does not switch between nematic and polar states either spatially or temporally.

Our experiments reveal that a small change in the degree of confinement can qualitatively alter the collective swimming behaviors of bacteria. The critical gap thickness of the Hele-Shaw cell, $h_c$, where the transition between the nematic phase and the cluster phase occurs, should be around twice the width of bacteria $w_d$ at $2w_d \approx 2$ {\textmu}m. However, the precise value of $h_c$ is hard to assess \textit{a priori} due to the natural variation in bacterial shapes and the complex interaction between bacteria and solid boundaries \cite{conrad2018confined}. Furthermore, as the gap thickness at such a small scale is hard to control accurately, experiments with seemingly similar geometries may result in completely different emergent phases. Thus, our results help to resolve the contradictory findings of previous experiments \cite{nishiguchi2017long,swiecicki2013swimming}. As the thickness of the Hele-Shaw cell is increased further from the quasi-2D geometry to the bulk limit, we would expect the emergence of bacterial turbulence in bulk suspensions, where the long-range hydrodynamic interaction overcomes the steric interaction and dictates the collective swimming of bacteria \cite{peng2021imaging}. The nature of the transition between the long-range nematic phase to bacterial turbulence upon increasing the gap thickness is an interesting direction for a future study. In general, the ability to control the emergent behaviors of active matter via geometric confinement provides not only a powerful method to probe the intrinsic dynamics of active matter but also a practical tool to tailor the behaviors of active matter in potential engineering applications.

\begin{acknowledgments}
We are indebted to Shuo Guo for helping with the experiments and for providing valuable insights. We also thank Renan Gross and Oliver Meacock for helping with the code used for segmenting and tracking bacteria in 2D, and thank Fernando Peruani and Robert Gro{\ss}mann for helpful discussions. This work was
funded by NSF CBET 1702352 and 2028652 and the David and Lucile Packard Foundation. We dedicate this paper to the memory of our mentor and friend, Prof. James W. Swan.
\end{acknowledgments}


\begin{thebibliography}{49}%
\makeatletter
\providecommand \@ifxundefined [1]{%
 \@ifx{#1\undefined}
}%
\providecommand \@ifnum [1]{%
 \ifnum #1\expandafter \@firstoftwo
 \else \expandafter \@secondoftwo
 \fi
}%
\providecommand \@ifx [1]{%
 \ifx #1\expandafter \@firstoftwo
 \else \expandafter \@secondoftwo
 \fi
}%
\providecommand \natexlab [1]{#1}%
\providecommand \enquote  [1]{``#1''}%
\providecommand \bibnamefont  [1]{#1}%
\providecommand \bibfnamefont [1]{#1}%
\providecommand \citenamefont [1]{#1}%
\providecommand \href@noop [0]{\@secondoftwo}%
\providecommand \href [0]{\begingroup \@sanitize@url \@href}%
\providecommand \@href[1]{\@@startlink{#1}\@@href}%
\providecommand \@@href[1]{\endgroup#1\@@endlink}%
\providecommand \@sanitize@url [0]{\catcode `\\12\catcode `\$12\catcode
  `\&12\catcode `\#12\catcode `\^12\catcode `\_12\catcode `\%12\relax}%
\providecommand \@@startlink[1]{}%
\providecommand \@@endlink[0]{}%
\providecommand \url  [0]{\begingroup\@sanitize@url \@url }%
\providecommand \@url [1]{\endgroup\@href {#1}{\urlprefix }}%
\providecommand \urlprefix  [0]{URL }%
\providecommand \Eprint [0]{\href }%
\providecommand \doibase [0]{https://doi.org/}%
\providecommand \selectlanguage [0]{\@gobble}%
\providecommand \bibinfo  [0]{\@secondoftwo}%
\providecommand \bibfield  [0]{\@secondoftwo}%
\providecommand \translation [1]{[#1]}%
\providecommand \BibitemOpen [0]{}%
\providecommand \bibitemStop [0]{}%
\providecommand \bibitemNoStop [0]{.\EOS\space}%
\providecommand \EOS [0]{\spacefactor3000\relax}%
\providecommand \BibitemShut  [1]{\csname bibitem#1\endcsname}%
\let\auto@bib@innerbib\@empty
\bibitem [{\citenamefont {Ramaswamy}(2010)}]{ramaswamy2010mechanics}%
  \BibitemOpen
  \bibfield  {author} {\bibinfo {author} {\bibfnamefont {S.}~\bibnamefont
  {Ramaswamy}},\ }\bibfield  {title} {\bibinfo {title} {The mechanics and
  statistics of active matter},\ }\href@noop {} {\bibfield  {journal} {\bibinfo
   {journal} {Annu. Rev. Condens. Matter Phys.}\ }\textbf {\bibinfo {volume}
  {1}},\ \bibinfo {pages} {323} (\bibinfo {year} {2010})}\BibitemShut {NoStop}%
\bibitem [{\citenamefont {Marchetti}\ \emph {et~al.}(2013)\citenamefont
  {Marchetti}, \citenamefont {Joanny}, \citenamefont {Ramaswamy}, \citenamefont
  {Liverpool}, \citenamefont {Prost}, \citenamefont {Rao},\ and\ \citenamefont
  {Simha}}]{marchetti2013hydrodynamics}%
  \BibitemOpen
  \bibfield  {author} {\bibinfo {author} {\bibfnamefont {M.~C.}\ \bibnamefont
  {Marchetti}}, \bibinfo {author} {\bibfnamefont {J.-F.}\ \bibnamefont
  {Joanny}}, \bibinfo {author} {\bibfnamefont {S.}~\bibnamefont {Ramaswamy}},
  \bibinfo {author} {\bibfnamefont {T.~B.}\ \bibnamefont {Liverpool}}, \bibinfo
  {author} {\bibfnamefont {J.}~\bibnamefont {Prost}}, \bibinfo {author}
  {\bibfnamefont {M.}~\bibnamefont {Rao}},\ and\ \bibinfo {author}
  {\bibfnamefont {R.~A.}\ \bibnamefont {Simha}},\ }\bibfield  {title} {\bibinfo
  {title} {Hydrodynamics of soft active matter},\ }\href@noop {} {\bibfield
  {journal} {\bibinfo  {journal} {Rev. Mod. Phys.}\ }\textbf {\bibinfo {volume}
  {85}},\ \bibinfo {pages} {1143} (\bibinfo {year} {2013})}\BibitemShut
  {NoStop}%
\bibitem [{\citenamefont {Schwarz-Linek}\ \emph {et~al.}(2016)\citenamefont
  {Schwarz-Linek}, \citenamefont {Arlt}, \citenamefont {Jepson}, \citenamefont
  {Dawson}, \citenamefont {Vissers}, \citenamefont {Miroli}, \citenamefont
  {Pilizota}, \citenamefont {Martinez},\ and\ \citenamefont
  {Poon}}]{schwarz2016review}%
  \BibitemOpen
  \bibfield  {author} {\bibinfo {author} {\bibfnamefont {J.}~\bibnamefont
  {Schwarz-Linek}}, \bibinfo {author} {\bibfnamefont {J.}~\bibnamefont {Arlt}},
  \bibinfo {author} {\bibfnamefont {A.}~\bibnamefont {Jepson}}, \bibinfo
  {author} {\bibfnamefont {A.}~\bibnamefont {Dawson}}, \bibinfo {author}
  {\bibfnamefont {T.}~\bibnamefont {Vissers}}, \bibinfo {author} {\bibfnamefont
  {D.}~\bibnamefont {Miroli}}, \bibinfo {author} {\bibfnamefont
  {T.}~\bibnamefont {Pilizota}}, \bibinfo {author} {\bibfnamefont {V.~A.}\
  \bibnamefont {Martinez}},\ and\ \bibinfo {author} {\bibfnamefont {W.~C.}\
  \bibnamefont {Poon}},\ }\bibfield  {title} {\bibinfo {title}
  {\emph{Escherichia coli} as a model active colloid: A practical
  introduction},\ }\href
  {https://www.sciencedirect.com/science/article/pii/S0927776515300825}
  {\bibfield  {journal} {\bibinfo  {journal} {Colloids Surf. B}\ }\textbf
  {\bibinfo {volume} {137}},\ \bibinfo {pages} {2} (\bibinfo {year}
  {2016})}\BibitemShut {NoStop}%
\bibitem [{\citenamefont {Be’er}\ and\ \citenamefont
  {Ariel}(2019)}]{beer2019review}%
  \BibitemOpen
  \bibfield  {author} {\bibinfo {author} {\bibfnamefont {A.}~\bibnamefont
  {Be’er}}\ and\ \bibinfo {author} {\bibfnamefont {G.}~\bibnamefont
  {Ariel}},\ }\bibfield  {title} {\bibinfo {title} {A statistical physics view
  of swarming bacteria},\ }\href@noop {} {\bibfield  {journal} {\bibinfo
  {journal} {Mov. Ecol.}\ }\textbf {\bibinfo {volume} {7}},\ \bibinfo {pages}
  {9} (\bibinfo {year} {2019})}\BibitemShut {NoStop}%
\bibitem [{\citenamefont {L\'opez}\ \emph {et~al.}(2015)\citenamefont
  {L\'opez}, \citenamefont {Gachelin}, \citenamefont {Douarche}, \citenamefont
  {Auradou},\ and\ \citenamefont {Cl\'ement}}]{lopez2015superfluids}%
  \BibitemOpen
  \bibfield  {author} {\bibinfo {author} {\bibfnamefont {H.~M.}\ \bibnamefont
  {L\'opez}}, \bibinfo {author} {\bibfnamefont {J.}~\bibnamefont {Gachelin}},
  \bibinfo {author} {\bibfnamefont {C.}~\bibnamefont {Douarche}}, \bibinfo
  {author} {\bibfnamefont {H.}~\bibnamefont {Auradou}},\ and\ \bibinfo {author}
  {\bibfnamefont {E.}~\bibnamefont {Cl\'ement}},\ }\bibfield  {title} {\bibinfo
  {title} {Turning bacteria suspensions into superfluids},\ }\href
  {https://link.aps.org/doi/10.1103/PhysRevLett.115.028301} {\bibfield
  {journal} {\bibinfo  {journal} {Phys. Rev. Lett.}\ }\textbf {\bibinfo
  {volume} {115}},\ \bibinfo {pages} {028301} (\bibinfo {year}
  {2015})}\BibitemShut {NoStop}%
\bibitem [{\citenamefont {Peng}\ \emph {et~al.}(2016)\citenamefont {Peng},
  \citenamefont {Lai}, \citenamefont {Tai}, \citenamefont {Zhang},
  \citenamefont {Xu},\ and\ \citenamefont {Cheng}}]{peng2016diffusion}%
  \BibitemOpen
  \bibfield  {author} {\bibinfo {author} {\bibfnamefont {Y.}~\bibnamefont
  {Peng}}, \bibinfo {author} {\bibfnamefont {L.}~\bibnamefont {Lai}}, \bibinfo
  {author} {\bibfnamefont {Y.-S.}\ \bibnamefont {Tai}}, \bibinfo {author}
  {\bibfnamefont {K.}~\bibnamefont {Zhang}}, \bibinfo {author} {\bibfnamefont
  {X.}~\bibnamefont {Xu}},\ and\ \bibinfo {author} {\bibfnamefont
  {X.}~\bibnamefont {Cheng}},\ }\bibfield  {title} {\bibinfo {title} {Diffusion
  of ellipsoids in bacterial suspensions},\ }\href@noop {} {\bibfield
  {journal} {\bibinfo  {journal} {Phys. Rev. Lett.}\ }\textbf {\bibinfo
  {volume} {116}},\ \bibinfo {pages} {068303} (\bibinfo {year}
  {2016})}\BibitemShut {NoStop}%
\bibitem [{\citenamefont {Guo}\ \emph {et~al.}(2018)\citenamefont {Guo},
  \citenamefont {Samanta}, \citenamefont {Peng}, \citenamefont {Xu},\ and\
  \citenamefont {Cheng}}]{guo2018shearbanding}%
  \BibitemOpen
  \bibfield  {author} {\bibinfo {author} {\bibfnamefont {S.}~\bibnamefont
  {Guo}}, \bibinfo {author} {\bibfnamefont {D.}~\bibnamefont {Samanta}},
  \bibinfo {author} {\bibfnamefont {Y.}~\bibnamefont {Peng}}, \bibinfo {author}
  {\bibfnamefont {X.}~\bibnamefont {Xu}},\ and\ \bibinfo {author}
  {\bibfnamefont {X.}~\bibnamefont {Cheng}},\ }\bibfield  {title} {\bibinfo
  {title} {Symmetric shear banding and swarming vortices in bacterial
  superfluids},\ }\href {https://www.pnas.org/content/115/28/7212} {\bibfield
  {journal} {\bibinfo  {journal} {Proc. Natl. Acad. Sci. U.S.A.}\ }\textbf
  {\bibinfo {volume} {115}},\ \bibinfo {pages} {7212} (\bibinfo {year}
  {2018})}\BibitemShut {NoStop}%
\bibitem [{\citenamefont {Conrad}\ and\ \citenamefont
  {Poling-Skutvik}(2018)}]{conrad2018confined}%
  \BibitemOpen
  \bibfield  {author} {\bibinfo {author} {\bibfnamefont {J.~C.}\ \bibnamefont
  {Conrad}}\ and\ \bibinfo {author} {\bibfnamefont {R.}~\bibnamefont
  {Poling-Skutvik}},\ }\bibfield  {title} {\bibinfo {title} {Confined flow:
  consequences and implications for bacteria and biofilms},\ }\href@noop {}
  {\bibfield  {journal} {\bibinfo  {journal} {Annu. Rev. Chem. Biomol. Eng.}\
  }\textbf {\bibinfo {volume} {9}},\ \bibinfo {pages} {175} (\bibinfo {year}
  {2018})}\BibitemShut {NoStop}%
\bibitem [{\citenamefont {Ranjard}\ and\ \citenamefont
  {Richaume}(2001)}]{ranjard2001quantitative}%
  \BibitemOpen
  \bibfield  {author} {\bibinfo {author} {\bibfnamefont {L.}~\bibnamefont
  {Ranjard}}\ and\ \bibinfo {author} {\bibfnamefont {A.}~\bibnamefont
  {Richaume}},\ }\bibfield  {title} {\bibinfo {title} {Quantitative and
  qualitative microscale distribution of bacteria in soil},\ }\href@noop {}
  {\bibfield  {journal} {\bibinfo  {journal} {Res. Microbiol.}\ }\textbf
  {\bibinfo {volume} {152}},\ \bibinfo {pages} {707} (\bibinfo {year}
  {2001})}\BibitemShut {NoStop}%
\bibitem [{\citenamefont {Bhattacharjee}\ and\ \citenamefont
  {Datta}(2019)}]{Bhattacharjee2019}%
  \BibitemOpen
  \bibfield  {author} {\bibinfo {author} {\bibfnamefont {T.}~\bibnamefont
  {Bhattacharjee}}\ and\ \bibinfo {author} {\bibfnamefont {S.~S.}\ \bibnamefont
  {Datta}},\ }\bibfield  {title} {\bibinfo {title} {Bacterial hopping and
  trapping in porous media},\ }\href@noop {} {\bibfield  {journal} {\bibinfo
  {journal} {Nat. Commun.}\ }\textbf {\bibinfo {volume} {10}},\ \bibinfo
  {pages} {2075} (\bibinfo {year} {2019})}\BibitemShut {NoStop}%
\bibitem [{\citenamefont {Ki}\ and\ \citenamefont
  {Rotstein}(2008)}]{ki2008bacterial}%
  \BibitemOpen
  \bibfield  {author} {\bibinfo {author} {\bibfnamefont {V.}~\bibnamefont
  {Ki}}\ and\ \bibinfo {author} {\bibfnamefont {C.}~\bibnamefont {Rotstein}},\
  }\bibfield  {title} {\bibinfo {title} {Bacterial skin and soft tissue
  infections in adults: a review of their epidemiology, pathogenesis,
  diagnosis, treatment and site of care},\ }\href@noop {} {\bibfield  {journal}
  {\bibinfo  {journal} {Can. J. Infect. Dis. Med. Microbiol.}\ }\textbf
  {\bibinfo {volume} {19}},\ \bibinfo {pages} {173} (\bibinfo {year}
  {2008})}\BibitemShut {NoStop}%
\bibitem [{\citenamefont {Dombrowski}\ \emph {et~al.}(2004)\citenamefont
  {Dombrowski}, \citenamefont {Cisneros}, \citenamefont {Chatkaew},
  \citenamefont {Goldstein},\ and\ \citenamefont
  {Kessler}}]{Dombrowski2004collective}%
  \BibitemOpen
  \bibfield  {author} {\bibinfo {author} {\bibfnamefont {C.}~\bibnamefont
  {Dombrowski}}, \bibinfo {author} {\bibfnamefont {L.}~\bibnamefont
  {Cisneros}}, \bibinfo {author} {\bibfnamefont {S.}~\bibnamefont {Chatkaew}},
  \bibinfo {author} {\bibfnamefont {R.~E.}\ \bibnamefont {Goldstein}},\ and\
  \bibinfo {author} {\bibfnamefont {J.~O.}\ \bibnamefont {Kessler}},\
  }\bibfield  {title} {\bibinfo {title} {Self-concentration and large-scale
  coherence in bacterial dynamics},\ }\href
  {https://link.aps.org/doi/10.1103/PhysRevLett.93.098103} {\bibfield
  {journal} {\bibinfo  {journal} {Phys. Rev. Lett.}\ }\textbf {\bibinfo
  {volume} {93}},\ \bibinfo {pages} {098103} (\bibinfo {year}
  {2004})}\BibitemShut {NoStop}%
\bibitem [{\citenamefont {Sokolov}\ \emph {et~al.}(2009)\citenamefont
  {Sokolov}, \citenamefont {Goldstein}, \citenamefont {Feldchtein},\ and\
  \citenamefont {Aranson}}]{Sokolov2009collective}%
  \BibitemOpen
  \bibfield  {author} {\bibinfo {author} {\bibfnamefont {A.}~\bibnamefont
  {Sokolov}}, \bibinfo {author} {\bibfnamefont {R.~E.}\ \bibnamefont
  {Goldstein}}, \bibinfo {author} {\bibfnamefont {F.~I.}\ \bibnamefont
  {Feldchtein}},\ and\ \bibinfo {author} {\bibfnamefont {I.~S.}\ \bibnamefont
  {Aranson}},\ }\bibfield  {title} {\bibinfo {title} {Enhanced mixing and
  spatial instability in concentrated bacterial suspensions},\ }\href
  {https://link.aps.org/doi/10.1103/PhysRevE.80.031903} {\bibfield  {journal}
  {\bibinfo  {journal} {Phys. Rev. E}\ }\textbf {\bibinfo {volume} {80}},\
  \bibinfo {pages} {031903} (\bibinfo {year} {2009})}\BibitemShut {NoStop}%
\bibitem [{\citenamefont {Wensink}\ \emph {et~al.}(2012)\citenamefont
  {Wensink}, \citenamefont {Dunkel}, \citenamefont {Heidenreich}, \citenamefont
  {Drescher}, \citenamefont {Goldstein}, \citenamefont {L{\"o}wen},\ and\
  \citenamefont {Yeomans}}]{wensink2012meso}%
  \BibitemOpen
  \bibfield  {author} {\bibinfo {author} {\bibfnamefont {H.~H.}\ \bibnamefont
  {Wensink}}, \bibinfo {author} {\bibfnamefont {J.}~\bibnamefont {Dunkel}},
  \bibinfo {author} {\bibfnamefont {S.}~\bibnamefont {Heidenreich}}, \bibinfo
  {author} {\bibfnamefont {K.}~\bibnamefont {Drescher}}, \bibinfo {author}
  {\bibfnamefont {R.~E.}\ \bibnamefont {Goldstein}}, \bibinfo {author}
  {\bibfnamefont {H.}~\bibnamefont {L{\"o}wen}},\ and\ \bibinfo {author}
  {\bibfnamefont {J.~M.}\ \bibnamefont {Yeomans}},\ }\bibfield  {title}
  {\bibinfo {title} {Meso-scale turbulence in living fluids},\ }\href@noop {}
  {\bibfield  {journal} {\bibinfo  {journal} {Proc. Natl. Acad. Sci. U.S.A.}\
  }\textbf {\bibinfo {volume} {109}},\ \bibinfo {pages} {14308} (\bibinfo
  {year} {2012})}\BibitemShut {NoStop}%
\bibitem [{\citenamefont {Peng}\ \emph {et~al.}(2021)\citenamefont {Peng},
  \citenamefont {Liu},\ and\ \citenamefont {Cheng}}]{peng2021imaging}%
  \BibitemOpen
  \bibfield  {author} {\bibinfo {author} {\bibfnamefont {Y.}~\bibnamefont
  {Peng}}, \bibinfo {author} {\bibfnamefont {Z.}~\bibnamefont {Liu}},\ and\
  \bibinfo {author} {\bibfnamefont {X.}~\bibnamefont {Cheng}},\ }\bibfield
  {title} {\bibinfo {title} {Imaging the emergence of bacterial turbulence:
  Phase diagram and transition kinetics},\ }\href@noop {} {\bibfield  {journal}
  {\bibinfo  {journal} {Sci. Adv.}\ }\textbf {\bibinfo {volume} {7}},\ \bibinfo
  {pages} {eabd1240} (\bibinfo {year} {2021})}\BibitemShut {NoStop}%
\bibitem [{\citenamefont {Liu}\ \emph {et~al.}()\citenamefont {Liu},
  \citenamefont {Zeng}, \citenamefont {Ma},\ and\ \citenamefont
  {Cheng}}]{liu2020density}%
  \BibitemOpen
  \bibfield  {author} {\bibinfo {author} {\bibfnamefont {Z.}~\bibnamefont
  {Liu}}, \bibinfo {author} {\bibfnamefont {W.}~\bibnamefont {Zeng}}, \bibinfo
  {author} {\bibfnamefont {X.}~\bibnamefont {Ma}},\ and\ \bibinfo {author}
  {\bibfnamefont {X.}~\bibnamefont {Cheng}},\ }\bibfield  {title} {\bibinfo
  {title} {Density fluctuations and energy spectra of 3{D} bacterial
  suspensions},\ }\href@noop {} {\bibinfo  {journal} {arXiv:2012.13680}\
  }\BibitemShut {NoStop}%
\bibitem [{\citenamefont {Brotto}\ \emph {et~al.}(2013)\citenamefont {Brotto},
  \citenamefont {Caussin}, \citenamefont {Lauga},\ and\ \citenamefont
  {Bartolo}}]{brotto2013hydrodynamics}%
  \BibitemOpen
\bibfield  {journal} {  }\bibfield  {author} {\bibinfo {author} {\bibfnamefont
  {T.}~\bibnamefont {Brotto}}, \bibinfo {author} {\bibfnamefont {J.-B.}\
  \bibnamefont {Caussin}}, \bibinfo {author} {\bibfnamefont {E.}~\bibnamefont
  {Lauga}},\ and\ \bibinfo {author} {\bibfnamefont {D.}~\bibnamefont
  {Bartolo}},\ }\bibfield  {title} {\bibinfo {title} {Hydrodynamics of confined
  active fluids},\ }\href@noop {} {\bibfield  {journal} {\bibinfo  {journal}
  {Phys. Rev. Lett.}\ }\textbf {\bibinfo {volume} {110}},\ \bibinfo {pages}
  {038101} (\bibinfo {year} {2013})}\BibitemShut {NoStop}%
\bibitem [{\citenamefont {Tsang}\ and\ \citenamefont
  {Kanso}(2014)}]{tsang2014flagella}%
  \BibitemOpen
  \bibfield  {author} {\bibinfo {author} {\bibfnamefont {A.~C.~H.}\
  \bibnamefont {Tsang}}\ and\ \bibinfo {author} {\bibfnamefont
  {E.}~\bibnamefont {Kanso}},\ }\bibfield  {title} {\bibinfo {title}
  {Flagella-induced transitions in the collective behavior of confined
  microswimmers},\ }\href@noop {} {\bibfield  {journal} {\bibinfo  {journal}
  {Phys. Rev. E}\ }\textbf {\bibinfo {volume} {90}},\ \bibinfo {pages} {021001}
  (\bibinfo {year} {2014})}\BibitemShut {NoStop}%
\bibitem [{\citenamefont {Aranson}\ \emph {et~al.}(2007)\citenamefont
  {Aranson}, \citenamefont {Sokolov}, \citenamefont {Kessler},\ and\
  \citenamefont {Goldstein}}]{Aranson2007steric}%
  \BibitemOpen
  \bibfield  {author} {\bibinfo {author} {\bibfnamefont {I.~S.}\ \bibnamefont
  {Aranson}}, \bibinfo {author} {\bibfnamefont {A.}~\bibnamefont {Sokolov}},
  \bibinfo {author} {\bibfnamefont {J.~O.}\ \bibnamefont {Kessler}},\ and\
  \bibinfo {author} {\bibfnamefont {R.~E.}\ \bibnamefont {Goldstein}},\
  }\bibfield  {title} {\bibinfo {title} {Model for dynamical coherence in thin
  films of self-propelled microorganisms},\ }\href
  {https://link.aps.org/doi/10.1103/PhysRevE.75.040901} {\bibfield  {journal}
  {\bibinfo  {journal} {Phys. Rev. E}\ }\textbf {\bibinfo {volume} {75}},\
  \bibinfo {pages} {040901} (\bibinfo {year} {2007})}\BibitemShut {NoStop}%
\bibitem [{\citenamefont {Baskaran}\ and\ \citenamefont
  {Marchetti}(2010)}]{baskaran2010steric}%
  \BibitemOpen
  \bibfield  {author} {\bibinfo {author} {\bibfnamefont {A.}~\bibnamefont
  {Baskaran}}\ and\ \bibinfo {author} {\bibfnamefont {M.~C.}\ \bibnamefont
  {Marchetti}},\ }\bibfield  {title} {\bibinfo {title} {Nonequilibrium
  statistical mechanics of self-propelled hard rods},\ }\href
  {https://doi.org/10.1088/1742-5468/2010/04/p04019} {\bibfield  {journal}
  {\bibinfo  {journal} {J. Stat. Mech. Theory Exp.}\ }\textbf {\bibinfo
  {volume} {2010}},\ \bibinfo {pages} {P04019} (\bibinfo {year}
  {2010})}\BibitemShut {NoStop}%
\bibitem [{\citenamefont {Nishiguchi}\ \emph {et~al.}(2017)\citenamefont
  {Nishiguchi}, \citenamefont {Nagai}, \citenamefont {Chat{\'e}},\ and\
  \citenamefont {Sano}}]{nishiguchi2017long}%
  \BibitemOpen
  \bibfield  {author} {\bibinfo {author} {\bibfnamefont {D.}~\bibnamefont
  {Nishiguchi}}, \bibinfo {author} {\bibfnamefont {K.~H.}\ \bibnamefont
  {Nagai}}, \bibinfo {author} {\bibfnamefont {H.}~\bibnamefont {Chat{\'e}}},\
  and\ \bibinfo {author} {\bibfnamefont {M.}~\bibnamefont {Sano}},\ }\bibfield
  {title} {\bibinfo {title} {Long-range nematic order and anomalous
  fluctuations in suspensions of swimming filamentous bacteria},\ }\href@noop
  {} {\bibfield  {journal} {\bibinfo  {journal} {Phys. Rev. E}\ }\textbf
  {\bibinfo {volume} {95}},\ \bibinfo {pages} {020601} (\bibinfo {year}
  {2017})}\BibitemShut {NoStop}%
\bibitem [{\citenamefont {Swiecicki}\ \emph {et~al.}(2013)\citenamefont
  {Swiecicki}, \citenamefont {Sliusarenko},\ and\ \citenamefont
  {Weibel}}]{swiecicki2013swimming}%
  \BibitemOpen
  \bibfield  {author} {\bibinfo {author} {\bibfnamefont {J.-M.}\ \bibnamefont
  {Swiecicki}}, \bibinfo {author} {\bibfnamefont {O.}~\bibnamefont
  {Sliusarenko}},\ and\ \bibinfo {author} {\bibfnamefont {D.~B.}\ \bibnamefont
  {Weibel}},\ }\bibfield  {title} {\bibinfo {title} {From swimming to swarming:
  \emph{Escherichia coli} cell motility in two-dimensions},\ }\href@noop {}
  {\bibfield  {journal} {\bibinfo  {journal} {Integr. Biol.}\ }\textbf
  {\bibinfo {volume} {5}},\ \bibinfo {pages} {1490} (\bibinfo {year}
  {2013})}\BibitemShut {NoStop}%
\bibitem [{\citenamefont {Be’er}\ \emph {et~al.}(2020)\citenamefont
  {Be’er}, \citenamefont {Ilkanaiv}, \citenamefont {Gross}, \citenamefont
  {Kearns}, \citenamefont {Heidenreich}, \citenamefont {B{\"a}r},\ and\
  \citenamefont {Ariel}}]{be2020phase}%
  \BibitemOpen
  \bibfield  {author} {\bibinfo {author} {\bibfnamefont {A.}~\bibnamefont
  {Be’er}}, \bibinfo {author} {\bibfnamefont {B.}~\bibnamefont {Ilkanaiv}},
  \bibinfo {author} {\bibfnamefont {R.}~\bibnamefont {Gross}}, \bibinfo
  {author} {\bibfnamefont {D.~B.}\ \bibnamefont {Kearns}}, \bibinfo {author}
  {\bibfnamefont {S.}~\bibnamefont {Heidenreich}}, \bibinfo {author}
  {\bibfnamefont {M.}~\bibnamefont {B{\"a}r}},\ and\ \bibinfo {author}
  {\bibfnamefont {G.}~\bibnamefont {Ariel}},\ }\bibfield  {title} {\bibinfo
  {title} {A phase diagram for bacterial swarming},\ }\href@noop {} {\bibfield
  {journal} {\bibinfo  {journal} {Commun. Phys.}\ }\textbf {\bibinfo {volume}
  {3}},\ \bibinfo {pages} {1} (\bibinfo {year} {2020})}\BibitemShut {NoStop}%
\bibitem [{\citenamefont {Wioland}\ \emph {et~al.}(2013)\citenamefont
  {Wioland}, \citenamefont {Woodhouse}, \citenamefont {Dunkel}, \citenamefont
  {Kessler},\ and\ \citenamefont {Goldstein}}]{wioland2013confinement}%
  \BibitemOpen
  \bibfield  {author} {\bibinfo {author} {\bibfnamefont {H.}~\bibnamefont
  {Wioland}}, \bibinfo {author} {\bibfnamefont {F.~G.}\ \bibnamefont
  {Woodhouse}}, \bibinfo {author} {\bibfnamefont {J.}~\bibnamefont {Dunkel}},
  \bibinfo {author} {\bibfnamefont {J.~O.}\ \bibnamefont {Kessler}},\ and\
  \bibinfo {author} {\bibfnamefont {R.~E.}\ \bibnamefont {Goldstein}},\
  }\bibfield  {title} {\bibinfo {title} {Confinement stabilizes a bacterial
  suspension into a spiral vortex},\ }\href@noop {} {\bibfield  {journal}
  {\bibinfo  {journal} {Phys. Rev. Lett.}\ }\textbf {\bibinfo {volume} {110}},\
  \bibinfo {pages} {268102} (\bibinfo {year} {2013})}\BibitemShut {NoStop}%
\bibitem [{\citenamefont {Lushi}\ \emph {et~al.}(2014)\citenamefont {Lushi},
  \citenamefont {Wioland},\ and\ \citenamefont
  {Goldstein}}]{lushi2014confinement}%
  \BibitemOpen
  \bibfield  {author} {\bibinfo {author} {\bibfnamefont {E.}~\bibnamefont
  {Lushi}}, \bibinfo {author} {\bibfnamefont {H.}~\bibnamefont {Wioland}},\
  and\ \bibinfo {author} {\bibfnamefont {R.~E.}\ \bibnamefont {Goldstein}},\
  }\bibfield  {title} {\bibinfo {title} {Fluid flows created by swimming
  bacteria drive self-organization in confined suspensions},\ }\href
  {https://www.pnas.org/content/111/27/9733} {\bibfield  {journal} {\bibinfo
  {journal} {Proc. Natl. Acad. Sci. U.S.A.}\ }\textbf {\bibinfo {volume}
  {111}},\ \bibinfo {pages} {9733} (\bibinfo {year} {2014})}\BibitemShut
  {NoStop}%
\bibitem [{\citenamefont {Wioland}\ \emph {et~al.}(2016)\citenamefont
  {Wioland}, \citenamefont {Lushi},\ and\ \citenamefont
  {Goldstein}}]{wioland2016confinement}%
  \BibitemOpen
  \bibfield  {author} {\bibinfo {author} {\bibfnamefont {H.}~\bibnamefont
  {Wioland}}, \bibinfo {author} {\bibfnamefont {E.}~\bibnamefont {Lushi}},\
  and\ \bibinfo {author} {\bibfnamefont {R.~E.}\ \bibnamefont {Goldstein}},\
  }\bibfield  {title} {\bibinfo {title} {Directed collective motion of bacteria
  under channel confinement},\ }\href
  {https://doi.org/10.1088/1367-2630/18/7/075002} {\bibfield  {journal}
  {\bibinfo  {journal} {New J. Phys.}\ }\textbf {\bibinfo {volume} {18}},\
  \bibinfo {pages} {075002} (\bibinfo {year} {2016})}\BibitemShut {NoStop}%
\bibitem [{\citenamefont {Liu}\ \emph {et~al.}(2019)\citenamefont {Liu},
  \citenamefont {Zhang},\ and\ \citenamefont {Cheng}}]{liu2019confinement}%
  \BibitemOpen
  \bibfield  {author} {\bibinfo {author} {\bibfnamefont {Z.}~\bibnamefont
  {Liu}}, \bibinfo {author} {\bibfnamefont {K.}~\bibnamefont {Zhang}},\ and\
  \bibinfo {author} {\bibfnamefont {X.}~\bibnamefont {Cheng}},\ }\bibfield
  {title} {\bibinfo {title} {Rheology of bacterial suspensions under
  confinement},\ }\href@noop {} {\bibfield  {journal} {\bibinfo  {journal}
  {Rheol. Acta}\ }\textbf {\bibinfo {volume} {58}},\ \bibinfo {pages} {439}
  (\bibinfo {year} {2019})}\BibitemShut {NoStop}%
\bibitem [{\citenamefont {Wu}(2015)}]{wu2015collective}%
  \BibitemOpen
  \bibfield  {author} {\bibinfo {author} {\bibfnamefont {Y.}~\bibnamefont
  {Wu}},\ }\bibfield  {title} {\bibinfo {title} {Collective motion of bacteria
  in two dimensions},\ }\href@noop {} {\bibfield  {journal} {\bibinfo
  {journal} {Quant. Biol.}\ }\textbf {\bibinfo {volume} {3}},\ \bibinfo {pages}
  {199} (\bibinfo {year} {2015})}\BibitemShut {NoStop}%
\bibitem [{\citenamefont {Kearns}(2010)}]{kearns2010field}%
  \BibitemOpen
  \bibfield  {author} {\bibinfo {author} {\bibfnamefont {D.~B.}\ \bibnamefont
  {Kearns}},\ }\bibfield  {title} {\bibinfo {title} {A field guide to bacterial
  swarming motility},\ }\href@noop {} {\bibfield  {journal} {\bibinfo
  {journal} {Nat. Rev. Microbiol.}\ }\textbf {\bibinfo {volume} {8}},\ \bibinfo
  {pages} {634} (\bibinfo {year} {2010})}\BibitemShut {NoStop}%
\bibitem [{\citenamefont {Zhang}\ \emph {et~al.}(2010)\citenamefont {Zhang},
  \citenamefont {Be’er}, \citenamefont {Florin},\ and\ \citenamefont
  {Swinney}}]{zhang2010collective}%
  \BibitemOpen
  \bibfield  {author} {\bibinfo {author} {\bibfnamefont {H.-P.}\ \bibnamefont
  {Zhang}}, \bibinfo {author} {\bibfnamefont {A.}~\bibnamefont {Be’er}},
  \bibinfo {author} {\bibfnamefont {E.-L.}\ \bibnamefont {Florin}},\ and\
  \bibinfo {author} {\bibfnamefont {H.~L.}\ \bibnamefont {Swinney}},\
  }\bibfield  {title} {\bibinfo {title} {Collective motion and density
  fluctuations in bacterial colonies},\ }\href@noop {} {\bibfield  {journal}
  {\bibinfo  {journal} {Proc. Natl. Acad. Sci. U.S.A.}\ }\textbf {\bibinfo
  {volume} {107}},\ \bibinfo {pages} {13626} (\bibinfo {year}
  {2010})}\BibitemShut {NoStop}%
\bibitem [{gho()}]{ghosh2021supplementary}%
  \BibitemOpen
  \href@noop {} {}\bibinfo {note} {See Supplemental Material at [URL will be
  inserted by publisher] for Movies 1 and 2.}\BibitemShut {Stop}%
\bibitem [{\citenamefont {Tanida}\ \emph {et~al.}(2020)\citenamefont {Tanida},
  \citenamefont {Furuta}, \citenamefont {Nishikawa}, \citenamefont {Hiraiwa},
  \citenamefont {Kojima}, \citenamefont {Oiwa},\ and\ \citenamefont
  {Sano}}]{tanida2020gliding}%
  \BibitemOpen
  \bibfield  {author} {\bibinfo {author} {\bibfnamefont {S.}~\bibnamefont
  {Tanida}}, \bibinfo {author} {\bibfnamefont {K.}~\bibnamefont {Furuta}},
  \bibinfo {author} {\bibfnamefont {K.}~\bibnamefont {Nishikawa}}, \bibinfo
  {author} {\bibfnamefont {T.}~\bibnamefont {Hiraiwa}}, \bibinfo {author}
  {\bibfnamefont {H.}~\bibnamefont {Kojima}}, \bibinfo {author} {\bibfnamefont
  {K.}~\bibnamefont {Oiwa}},\ and\ \bibinfo {author} {\bibfnamefont
  {M.}~\bibnamefont {Sano}},\ }\bibfield  {title} {\bibinfo {title} {Gliding
  filament system giving both global orientational order and clusters in
  collective motion},\ }\href@noop {} {\bibfield  {journal} {\bibinfo
  {journal} {Phys. Rev. E}\ }\textbf {\bibinfo {volume} {101}},\ \bibinfo
  {pages} {032607} (\bibinfo {year} {2020})}\BibitemShut {NoStop}%
\bibitem [{\citenamefont {Sciortino}\ and\ \citenamefont
  {Bausch}(2021)}]{sciortino2021pattern}%
  \BibitemOpen
  \bibfield  {author} {\bibinfo {author} {\bibfnamefont {A.}~\bibnamefont
  {Sciortino}}\ and\ \bibinfo {author} {\bibfnamefont {A.~R.}\ \bibnamefont
  {Bausch}},\ }\bibfield  {title} {\bibinfo {title} {Pattern formation and
  polarity sorting of driven actin filaments on lipid membranes},\ }\href@noop
  {} {\bibfield  {journal} {\bibinfo  {journal} {Proc. Natl. Acad. Sci.
  U.S.A.}\ }\textbf {\bibinfo {volume} {118}} (\bibinfo {year}
  {2021})}\BibitemShut {NoStop}%
\bibitem [{\citenamefont {Gro{\ss}mann}\ \emph {et~al.}(2020)\citenamefont
  {Gro{\ss}mann}, \citenamefont {Aranson},\ and\ \citenamefont
  {Peruani}}]{grossmann2020particle}%
  \BibitemOpen
  \bibfield  {author} {\bibinfo {author} {\bibfnamefont {R.}~\bibnamefont
  {Gro{\ss}mann}}, \bibinfo {author} {\bibfnamefont {I.~S.}\ \bibnamefont
  {Aranson}},\ and\ \bibinfo {author} {\bibfnamefont {F.}~\bibnamefont
  {Peruani}},\ }\bibfield  {title} {\bibinfo {title} {A particle-field approach
  bridges phase separation and collective motion in active matter},\
  }\href@noop {} {\bibfield  {journal} {\bibinfo  {journal} {Nat. Commun.}\
  }\textbf {\bibinfo {volume} {11}},\ \bibinfo {pages} {1} (\bibinfo {year}
  {2020})}\BibitemShut {NoStop}%
\bibitem [{\citenamefont {Denk}\ and\ \citenamefont
  {Frey}(2020)}]{denk2020pattern}%
  \BibitemOpen
  \bibfield  {author} {\bibinfo {author} {\bibfnamefont {J.}~\bibnamefont
  {Denk}}\ and\ \bibinfo {author} {\bibfnamefont {E.}~\bibnamefont {Frey}},\
  }\bibfield  {title} {\bibinfo {title} {Pattern-induced local symmetry
  breaking in active-matter systems},\ }\href@noop {} {\bibfield  {journal}
  {\bibinfo  {journal} {Proc. Natl. Acad. Sci. U.S.A.}\ }\textbf {\bibinfo
  {volume} {117}},\ \bibinfo {pages} {31623} (\bibinfo {year}
  {2020})}\BibitemShut {NoStop}%
\bibitem [{\citenamefont {B{\"a}r}\ \emph {et~al.}(2020)\citenamefont
  {B{\"a}r}, \citenamefont {Gro{\ss}mann}, \citenamefont {Heidenreich},\ and\
  \citenamefont {Peruani}}]{bar2020self}%
  \BibitemOpen
  \bibfield  {author} {\bibinfo {author} {\bibfnamefont {M.}~\bibnamefont
  {B{\"a}r}}, \bibinfo {author} {\bibfnamefont {R.}~\bibnamefont
  {Gro{\ss}mann}}, \bibinfo {author} {\bibfnamefont {S.}~\bibnamefont
  {Heidenreich}},\ and\ \bibinfo {author} {\bibfnamefont {F.}~\bibnamefont
  {Peruani}},\ }\bibfield  {title} {\bibinfo {title} {Self-propelled rods:
  Insights and perspectives for active matter},\ }\href@noop {} {\bibfield
  {journal} {\bibinfo  {journal} {Annu. Rev. Condens. Matter Phys.}\ }\textbf
  {\bibinfo {volume} {11}},\ \bibinfo {pages} {441} (\bibinfo {year}
  {2020})}\BibitemShut {NoStop}%
\bibitem [{\citenamefont {Ginelli}\ \emph {et~al.}(2010)\citenamefont
  {Ginelli}, \citenamefont {Peruani}, \citenamefont {B{\"a}r},\ and\
  \citenamefont {Chat{\'e}}}]{ginelli2010large}%
  \BibitemOpen
  \bibfield  {author} {\bibinfo {author} {\bibfnamefont {F.}~\bibnamefont
  {Ginelli}}, \bibinfo {author} {\bibfnamefont {F.}~\bibnamefont {Peruani}},
  \bibinfo {author} {\bibfnamefont {M.}~\bibnamefont {B{\"a}r}},\ and\ \bibinfo
  {author} {\bibfnamefont {H.}~\bibnamefont {Chat{\'e}}},\ }\bibfield  {title}
  {\bibinfo {title} {Large-scale collective properties of self-propelled
  rods},\ }\href@noop {} {\bibfield  {journal} {\bibinfo  {journal} {Phys. Rev.
  Lett.}\ }\textbf {\bibinfo {volume} {104}},\ \bibinfo {pages} {184502}
  (\bibinfo {year} {2010})}\BibitemShut {NoStop}%
\bibitem [{\citenamefont {Shi}\ and\ \citenamefont
  {Chat{\'e}}()}]{shi2018self}%
  \BibitemOpen
  \bibfield  {author} {\bibinfo {author} {\bibfnamefont {X.-q.}\ \bibnamefont
  {Shi}}\ and\ \bibinfo {author} {\bibfnamefont {H.}~\bibnamefont
  {Chat{\'e}}},\ }\bibfield  {title} {\bibinfo {title} {Self-propelled rods:
  Linking alignment-dominated and repulsion-dominated active matter},\
  }\href@noop {} {\bibinfo  {journal} {arXiv:1807.00294}\ }\BibitemShut
  {NoStop}%
\bibitem [{\citenamefont {Peruani}\ \emph {et~al.}(2006)\citenamefont
  {Peruani}, \citenamefont {Deutsch},\ and\ \citenamefont
  {B{\"a}r}}]{peruani2006nonequilibrium}%
  \BibitemOpen
\bibfield  {journal} {  }\bibfield  {author} {\bibinfo {author} {\bibfnamefont
  {F.}~\bibnamefont {Peruani}}, \bibinfo {author} {\bibfnamefont
  {A.}~\bibnamefont {Deutsch}},\ and\ \bibinfo {author} {\bibfnamefont
  {M.}~\bibnamefont {B{\"a}r}},\ }\bibfield  {title} {\bibinfo {title}
  {Nonequilibrium clustering of self-propelled rods},\ }\href@noop {}
  {\bibfield  {journal} {\bibinfo  {journal} {Phys. Rev. E}\ }\textbf {\bibinfo
  {volume} {74}},\ \bibinfo {pages} {030904} (\bibinfo {year}
  {2006})}\BibitemShut {NoStop}%
\bibitem [{\citenamefont {Abkenar}\ \emph {et~al.}(2013)\citenamefont
  {Abkenar}, \citenamefont {Marx}, \citenamefont {Auth},\ and\ \citenamefont
  {Gompper}}]{abkenar2013collective}%
  \BibitemOpen
  \bibfield  {author} {\bibinfo {author} {\bibfnamefont {M.}~\bibnamefont
  {Abkenar}}, \bibinfo {author} {\bibfnamefont {K.}~\bibnamefont {Marx}},
  \bibinfo {author} {\bibfnamefont {T.}~\bibnamefont {Auth}},\ and\ \bibinfo
  {author} {\bibfnamefont {G.}~\bibnamefont {Gompper}},\ }\bibfield  {title}
  {\bibinfo {title} {Collective behavior of penetrable self-propelled rods in
  two dimensions},\ }\href@noop {} {\bibfield  {journal} {\bibinfo  {journal}
  {Phys. Rev. E}\ }\textbf {\bibinfo {volume} {88}},\ \bibinfo {pages} {062314}
  (\bibinfo {year} {2013})}\BibitemShut {NoStop}%
\bibitem [{\citenamefont {Weitz}\ \emph {et~al.}(2015)\citenamefont {Weitz},
  \citenamefont {Deutsch},\ and\ \citenamefont {Peruani}}]{weitz2015self}%
  \BibitemOpen
  \bibfield  {author} {\bibinfo {author} {\bibfnamefont {S.}~\bibnamefont
  {Weitz}}, \bibinfo {author} {\bibfnamefont {A.}~\bibnamefont {Deutsch}},\
  and\ \bibinfo {author} {\bibfnamefont {F.}~\bibnamefont {Peruani}},\
  }\bibfield  {title} {\bibinfo {title} {Self-propelled rods exhibit a
  phase-separated state characterized by the presence of active stresses and
  the ejection of polar clusters},\ }\href@noop {} {\bibfield  {journal}
  {\bibinfo  {journal} {Phys. Rev. E}\ }\textbf {\bibinfo {volume} {92}},\
  \bibinfo {pages} {012322} (\bibinfo {year} {2015})}\BibitemShut {NoStop}%
\bibitem [{\citenamefont {Suzuki}\ \emph {et~al.}(2015)\citenamefont {Suzuki},
  \citenamefont {Weber}, \citenamefont {Frey},\ and\ \citenamefont
  {Bausch}}]{suzuki2015polar}%
  \BibitemOpen
  \bibfield  {author} {\bibinfo {author} {\bibfnamefont {R.}~\bibnamefont
  {Suzuki}}, \bibinfo {author} {\bibfnamefont {C.~A.}\ \bibnamefont {Weber}},
  \bibinfo {author} {\bibfnamefont {E.}~\bibnamefont {Frey}},\ and\ \bibinfo
  {author} {\bibfnamefont {A.~R.}\ \bibnamefont {Bausch}},\ }\bibfield  {title}
  {\bibinfo {title} {Polar pattern formation in driven filament systems
  requires non-binary particle collisions},\ }\href@noop {} {\bibfield
  {journal} {\bibinfo  {journal} {Nat. Phys.}\ }\textbf {\bibinfo {volume}
  {11}},\ \bibinfo {pages} {839} (\bibinfo {year} {2015})}\BibitemShut
  {NoStop}%
\bibitem [{\citenamefont {Yang}\ \emph {et~al.}(2010)\citenamefont {Yang},
  \citenamefont {Marceau},\ and\ \citenamefont {Gompper}}]{yang2010swarm}%
  \BibitemOpen
  \bibfield  {author} {\bibinfo {author} {\bibfnamefont {Y.}~\bibnamefont
  {Yang}}, \bibinfo {author} {\bibfnamefont {V.}~\bibnamefont {Marceau}},\ and\
  \bibinfo {author} {\bibfnamefont {G.}~\bibnamefont {Gompper}},\ }\bibfield
  {title} {\bibinfo {title} {Swarm behavior of self-propelled rods and swimming
  flagella},\ }\href@noop {} {\bibfield  {journal} {\bibinfo  {journal} {Phys.
  Rev. E}\ }\textbf {\bibinfo {volume} {82}},\ \bibinfo {pages} {031904}
  (\bibinfo {year} {2010})}\BibitemShut {NoStop}%
\bibitem [{\citenamefont {Vizsnyiczai}\ \emph {et~al.}(2020)\citenamefont
  {Vizsnyiczai}, \citenamefont {Frangipane}, \citenamefont {Bianchi},
  \citenamefont {Saglimbeni}, \citenamefont {Dell’Arciprete},\ and\
  \citenamefont {Di~Leonardo}}]{vizsnyiczai2020transition}%
  \BibitemOpen
  \bibfield  {author} {\bibinfo {author} {\bibfnamefont {G.}~\bibnamefont
  {Vizsnyiczai}}, \bibinfo {author} {\bibfnamefont {G.}~\bibnamefont
  {Frangipane}}, \bibinfo {author} {\bibfnamefont {S.}~\bibnamefont {Bianchi}},
  \bibinfo {author} {\bibfnamefont {F.}~\bibnamefont {Saglimbeni}}, \bibinfo
  {author} {\bibfnamefont {D.}~\bibnamefont {Dell’Arciprete}},\ and\ \bibinfo
  {author} {\bibfnamefont {R.}~\bibnamefont {Di~Leonardo}},\ }\bibfield
  {title} {\bibinfo {title} {A transition to stable one-dimensional swimming
  enhances \emph{E. coli} motility through narrow channels},\ }\href@noop {}
  {\bibfield  {journal} {\bibinfo  {journal} {Nat. Commun.}\ }\textbf {\bibinfo
  {volume} {11}},\ \bibinfo {pages} {1} (\bibinfo {year} {2020})}\BibitemShut
  {NoStop}%
\bibitem [{\citenamefont {Suzuki}\ and\ \citenamefont
  {Bausch}(2017)}]{suzuki2017emergence}%
  \BibitemOpen
  \bibfield  {author} {\bibinfo {author} {\bibfnamefont {R.}~\bibnamefont
  {Suzuki}}\ and\ \bibinfo {author} {\bibfnamefont {A.~R.}\ \bibnamefont
  {Bausch}},\ }\bibfield  {title} {\bibinfo {title} {The emergence and
  transient behaviour of collective motion in active filament systems},\
  }\href@noop {} {\bibfield  {journal} {\bibinfo  {journal} {Nat. Commun.}\
  }\textbf {\bibinfo {volume} {8}},\ \bibinfo {pages} {1} (\bibinfo {year}
  {2017})}\BibitemShut {NoStop}%
\bibitem [{\citenamefont {Huber}\ \emph {et~al.}(2018)\citenamefont {Huber},
  \citenamefont {Suzuki}, \citenamefont {Kr{\"u}ger}, \citenamefont {Frey},\
  and\ \citenamefont {Bausch}}]{huber2018emergence}%
  \BibitemOpen
  \bibfield  {author} {\bibinfo {author} {\bibfnamefont {L.}~\bibnamefont
  {Huber}}, \bibinfo {author} {\bibfnamefont {R.}~\bibnamefont {Suzuki}},
  \bibinfo {author} {\bibfnamefont {T.}~\bibnamefont {Kr{\"u}ger}}, \bibinfo
  {author} {\bibfnamefont {E.}~\bibnamefont {Frey}},\ and\ \bibinfo {author}
  {\bibfnamefont {A.}~\bibnamefont {Bausch}},\ }\bibfield  {title} {\bibinfo
  {title} {Emergence of coexisting ordered states in active matter systems},\
  }\href@noop {} {\bibfield  {journal} {\bibinfo  {journal} {Science}\ }\textbf
  {\bibinfo {volume} {361}},\ \bibinfo {pages} {255} (\bibinfo {year}
  {2018})}\BibitemShut {NoStop}%
\bibitem [{\citenamefont {Walter}\ \emph {et~al.}(2007)\citenamefont {Walter},
  \citenamefont {Greenfield}, \citenamefont {Bustamante},\ and\ \citenamefont
  {Liphardt}}]{Walter2007}%
  \BibitemOpen
  \bibfield  {author} {\bibinfo {author} {\bibfnamefont {J.~M.}\ \bibnamefont
  {Walter}}, \bibinfo {author} {\bibfnamefont {D.}~\bibnamefont {Greenfield}},
  \bibinfo {author} {\bibfnamefont {C.}~\bibnamefont {Bustamante}},\ and\
  \bibinfo {author} {\bibfnamefont {J.}~\bibnamefont {Liphardt}},\ }\bibfield
  {title} {\bibinfo {title} {Light-powering \emph{Escherichia coli} with
  proteorhodopsin},\ }\href {https://www.pnas.org/content/104/7/2408}
  {\bibfield  {journal} {\bibinfo  {journal} {Proc. Natl. Acad. Sci. U.S.A.}\
  }\textbf {\bibinfo {volume} {104}},\ \bibinfo {pages} {2408} (\bibinfo {year}
  {2007})}\BibitemShut {NoStop}%
\bibitem [{\citenamefont {Meacock}\ \emph {et~al.}(2021)\citenamefont
  {Meacock}, \citenamefont {Doostmohammadi}, \citenamefont {Foster},
  \citenamefont {Yeomans},\ and\ \citenamefont {Durham}}]{meacock2021bacteria}%
  \BibitemOpen
  \bibfield  {author} {\bibinfo {author} {\bibfnamefont {O.~J.}\ \bibnamefont
  {Meacock}}, \bibinfo {author} {\bibfnamefont {A.}~\bibnamefont
  {Doostmohammadi}}, \bibinfo {author} {\bibfnamefont {K.~R.}\ \bibnamefont
  {Foster}}, \bibinfo {author} {\bibfnamefont {J.~M.}\ \bibnamefont
  {Yeomans}},\ and\ \bibinfo {author} {\bibfnamefont {W.~M.}\ \bibnamefont
  {Durham}},\ }\bibfield  {title} {\bibinfo {title} {Bacteria solve the problem
  of crowding by moving slowly},\ }\href@noop {} {\bibfield  {journal}
  {\bibinfo  {journal} {Nat. Phys.}\ }\textbf {\bibinfo {volume} {17}},\
  \bibinfo {pages} {205} (\bibinfo {year} {2021})}\BibitemShut {NoStop}%
\bibitem [{\citenamefont {Rabani}\ \emph {et~al.}(2013)\citenamefont {Rabani},
  \citenamefont {Ariel},\ and\ \citenamefont {Be'er}}]{rabani2013collective}%
  \BibitemOpen
  \bibfield  {author} {\bibinfo {author} {\bibfnamefont {A.}~\bibnamefont
  {Rabani}}, \bibinfo {author} {\bibfnamefont {G.}~\bibnamefont {Ariel}},\ and\
  \bibinfo {author} {\bibfnamefont {A.}~\bibnamefont {Be'er}},\ }\bibfield
  {title} {\bibinfo {title} {Collective motion of spherical bacteria},\
  }\href@noop {} {\bibfield  {journal} {\bibinfo  {journal} {PloS one}\
  }\textbf {\bibinfo {volume} {8}},\ \bibinfo {pages} {e83760} (\bibinfo {year}
  {2013})}\BibitemShut {NoStop}%
\end{thebibliography}

\providecommand{\noopsort}[1]{}\providecommand{\singleletter}[1]{#1}%
%


\appendix

\section{\label{app:methods}Methods}
\subsection{Bacterial strain and culturing}


Wild-type \textit{E. coli} (BW25113) are genetically modified to express the transmembrane proton pump proteorhodopsin (PR), using the plasmid pZE-PR encoding the SAR86 $\gamma$-proteobacterial PR-variant \cite{Walter2007}. As the activity of PR is correlated with the intensity of incident light, we are able to control the swimming velocity of bacteria in the range of 4 {\textmu}m/s to 15 {\textmu}m/s.

To prepare a suspension of motile bacteria, we inoculate a small amount of bacterial frozen stock in 2 mL Terrific Broth (tryptone 1.2\% (w/v), yeast extract 2.4\% (w/v) and glycerol 0.4\% (w/v)). Bacteria are then incubated at 37 $^{\circ}$C for 15 hours in an orbital shaker operating at 250 rpm. This bacterial culture is then diluted 1:100 with fresh Terrific Broth and grown at 30 $^{\circ}$C for 6.5 hours. We add 1 mM isopropyl $\beta$-d-1-thiogalactopyranoside and 10 {\textmu}M methanolic all-trans-retinal in the mid-log phase of bacterial growth, to trigger the expression of PR. Finally, in the late log phase, bacteria are harvested by gentle centrifugation. The supernatant is discarded, and the bacteria are then resuspended in de-ionized water. The suspension is further washed twice and adjusted by adding water to reach the desired concentration.

\subsection{Hele-Shaw cell}

To create a Hele-Shaw geometry, we first deposit a droplet of \textit{E. coli} suspension of controlled volume on a glass slide. The droplet is then confined by gently pressing a glass coverslip of dimensions 18 mm by 18 mm onto it, ensuring the complete absence of air bubbles in the confined droplet. The edges of the coverslip are then sealed with UV-curable adhesive. To prevent bacteria from sticking on glass surfaces, glass slides and coverslips are treated in 1 M NaOH for 6 hours before use.  

The thickness of the Hele-Shaw cell is controlled by changing the volume of the confined droplet: a 0.7 {\textmu}L droplet is used for the 2D geometry and a 0.9 {\textmu}L droplet is used for the quasi-2D geometry.  As the gap thickness cannot be accurately measured, we approximate it by dividing the volume of the liquid droplet by the area of the glass coverslip  of 18 mm by 18 mm. For the quasi-2D geometry, this yields a gap thickness of 2.8 {\textmu}m, whereas, for the 2D geometry, the gap thickness is 2.2 {\textmu}m. More importantly, by direct visual inspection, we confirm that bacteria are able to cross over in the quasi-2D geometry but unable to do so in the 2D geometry.

\subsection{Video microscopy}

The dynamics of bacteria in Hele-Shaw cells are imaged through an inverted bright-field microscope (Nikon Ti-Eclipse) using a $60\times$ objective lens with a numerical aperture of 1.25. The field of view is set at 2320 pixels by 2080 pixels, which corresponds to a physical dimension of 232 {\textmu}m by 208 {\textmu}m. The swimming velocity of bacteria is controlled by changing the light intensity of the illumination lamp of the bright-field microscope. We first prepare Hele-Shaw cells containing different concentrations of bacteria. The light intensity is then varied to adjust the swimming speed of bacteria. The resultant emergent phase behavior is finally imaged. Videos are recorded at a frame rate of 30 frames per second for a total of 1000 frames using a scientific complementary metal-oxide-semiconductor (sCMOS) camera (Andor Zyla 4.2).

\subsection{Image processing and analysis}

The acquired videos are first preprocessed using custom-written MATLAB and Python scripts to remove background noise. The images are then binarized using Otsu’s method, where bacterial cells appear as white blobs in the image. The software package developed by Be'er \emph{et al.} \cite{be2020phase} is used to segment the bacterial cells. The area fraction occupied by bacteria in an image is defined as the ratio of the number of white pixels to the total number of pixels. The positions $\mathbf{r}$ and the orientations $\theta$ of these white blobs in the image, representing bacteria, are then extracted. 

In the 2D geometry, bacteria are unable to overlap or cross over each other. Hence, all the bacteria in the field of view can be accurately tracked. To determine the instantaneous velocities of bacteria in the 2D geometry, the FAST module developed by Meacock \emph{et al.} \cite{meacock2021bacteria} is used. The velocities thus obtained lie within 10\% of that measured by direct manual tracking. In the quasi-2D geometry, bacteria can cross over each other, making it difficult to track them from one frame to the next. To determine their instantaneous velocities in the quasi-2D geometry, we use the principle of optical flow \cite{rabani2013collective}. Particularly, we use the Farneback dense optical flow method to compute a velocity field for all pixels in the image. Then, we consider the velocity vectors contained in the white pixels in the binary image, representing bacterial bodies. To remove noise, we consider the range between 75\% and 90\% of these velocities and calculate their mean, which is used as the velocity of a bacterium between two frames. The velocities obtained using this definition again lie within 10\% of the velocities obtained using direct manual tracking, hence validating the method of optical flow. 

For each frame in the videos, we determine the area fraction, the mean velocity of bacteria and the nematic order parameter in case of the quasi-2D geometry, and the number of clusters in case of the 2D geometry. These values are then averaged over the 1000 frames of a video to give $\phi$, $V$, $S$, and $N_c$ reported in the main text.

To obtain long-time statistics of trajectories and calibrate our tracking algorithms, we also manually track bacteria for both the 2D and quasi-2D geometries at $\phi = 0.15$ and $V = 12$ {\textmu}m/s at a frame rate of 10 frames per second for 6 seconds. For the nematic phase, we track the motion of a total of 100 bacteria, whose trajectories are shown in Fig.~\ref{fig:fig1}(b). For the cluster phase, we first identify 17 different bacterial clusters and then track a total of 117 bacteria in these clusters. The bacterial trajectories are shown in Fig.~\ref{fig:fig1}(e). The trajectories of the bacteria in 2D are used to calculate the time dependence of the adjacent angles and pairwise distances in clusters (Fig.~\ref{fig:fig3}), and the neighbor dependence of bacterial velocities (Fig.~\ref{fig:fig4}). Trajectories from both geometries are used to calculate the directional persistence (Fig.~\ref{fig:fig5}(a)) and the velocity fluctuations (Fig.~\ref{fig:fig5}(b)).

\end{document}